\documentclass[fleqn,usenatbib]{mnras}

\usepackage[T1]{fontenc}

\DeclareRobustCommand{\VAN}[3]{#2}
\let\VANthebibliography\thebibliography
\def\thebibliography{\DeclareRobustCommand{\VAN}[3]{##3}\VANthebibliography}

\usepackage{graphicx}	
\graphicspath{{figures/}}
\usepackage{amsmath}	
\usepackage{amssymb}	
\usepackage{comment}
\usepackage{siunitx}
\usepackage{booktabs}
\usepackage{subcaption}
\usepackage[flushleft]{threeparttable}
\usepackage[ruled,vlined]{algorithm2e}
\usepackage{relsize}
\DeclareSIUnit \parsec {pc}
\usepackage{newtxtext,newtxmath}

\title[Coherent Search across all Keplerian parameters]{Coherent Search for Binary Pulsars across all Five Keplerian Parameters in Radio Observations using the template-bank algorithm}

\author[V.Balakrishnan et al.]{
Vishnu Balakrishnan,$^{1}$\thanks{E-mail: vishnu@mpifr-bonn.mpg.de}
David Champion,$^{1}$ Ewan Barr,$^{1}$ Michael Kramer,$^{1}$, V. Venkatraman Krishnan,$^{1}$,
\newauthor{}
Ralph P. Eatough,$^{1,2}$, Rahul Sengar,$^{3}$, Matthew Bailes$^{3}$
\\
$^{1}$Max-Planck-Institut fur Radioastronomie, Auf dem Hügel 69, D-53121 Bonn, Germany\\
$^{2}$National Astronomical Observatories, Chinese Academy of Sciences, 20A Datun Road, Chaoyang District, Beijing 100101, PR China \\
$^{3}$Centre for Astrophysics and Supercomputing, Swinburne University of Technology, P.O. Box 218, Hawthorn, Victoria 3122, Australia
}

\date{Accepted XXX. Received YYY; in original form ZZZ}

\pubyear{2015}

\begin{document}
\label{firstpage}
\pagerange{\pageref{firstpage}--\pageref{lastpage}}
\maketitle

\begin{abstract}
Relativistic binary pulsars orbiting white-dwarfs and neutron stars have already provided excellent tests of gravity. However, despite observational efforts, a pulsar orbiting a black hole has remained elusive. One possible explanation is the extreme Doppler smearing caused by the pulsar’s orbital motion which changes its apparent spin frequency during an observation. The classical solution to this problem has been to assume a constant acceleration/jerk for the entire observation. However, this assumption breaks down when the observation samples a large fraction of the orbit. This limits the length of search observations, and hence their sensitivity. This provides a strong motivation to develop techniques that can find compact binaries in longer observations. Here we present a GPU-based radio pulsar search pipeline that can perform a coherent search for binary pulsars by directly searching over three or five Keplerian parameters using the template-bank algorithm. We compare the sensitivity obtained from our pipeline with acceleration and jerk search pipelines for simulated pulsar-stellar-mass black hole binaries and observations of PSR~J0737-3039A. We also discuss the computational feasibility of our pipeline for untargeted pulsar surveys and targeted searches. Our benchmarks indicate that circular orbit searches for P-BH binaries with spin-period $P{_\mathrm{{spin}}} \geq \SI{20} {\milli \second}$ covering the 3-10~T$\mathrm{_{obs}}$ regime are feasible for the High Time Resolution Universe pulsar survey. Additionally, an elliptical orbit search in Globular clusters for $P{_\mathrm{{spin}}} \geq \SI{20} {\milli \second}$ pulsars orbiting intermediate-mass black holes  in the 5-10~T$\mathrm{_{obs}}$ regime is feasible for observations shorter than 2 hours with an eccentricity limit of 0.1.
\end{abstract}




\section{Introduction}
The discovery of the first binary pulsar PSR~B1913+16 and the subsequent measurement of its orbital period decay provided a new and exciting tool for testing theories of gravity in the strong field limit \citep{1975ApJ...195L..51H, 1982ApJ...253..908T, 1984PhRvL..52.1348W,2016ApJ...829...55W}. Another major milestone was the discovery of the first, and to date only, known double pulsar system PSR~J0737-3039A/B \citep{2003Natur.426..531B, 2004Sci...303.1153L}. Subsequent timing observations of this pulsar enabled measurement of seven Post-Keplerian parameters which provided multiple stringent tests of Einstein's theory of General Relativity (GR) \citep{2006Sci...314...97K, kramer21}. Additionally, timing observations of the first pulsar triple system PSR~J0337+1715 \citep{2018Natur.559...73A, 2020A&A...638A..24V} set a new lower limit for the parameter $\Delta$ describing a possible violation of the Strong Equivalence Principle (SEP), consistent with GR and tightly constraining multiple scalar-tensor theories of gravity. These examples are only some of the highlights that demonstrate the value of discovering new relativistic binary pulsars. The anticipated discovery of a pulsar orbiting a black hole will provide a unique tool for probing strong-field gravity and black hole physics.  \citet{1999ApJ...514..388W, 2014MNRAS.445.3115L} gave a detailed account of how the properties of the black hole can be studied by timing a pulsar orbiting the black hole. For example, the mass of a black hole can be uniquely determined by measuring two Post-Keplerian parameters. Additionally, the spin of a black hole can be measured by studying the precession of the pulsar orbit with time caused by relativistic spin-orbit coupling. Similarly, the quadrupole moment of the black hole's external gravitational field can be detected with upcoming facilities like the Square Kilometre Array (SKA) \citep{2004astro.ph..9020K}. \citet{2014MNRAS.445.3115L} demonstrated that by timing a pulsar orbiting a stellar-mass black hole (SBH) for few years would allow precise measurements of the black hole mass and spin. Such a discovery would help us test the Cosmic Censorship Conjecture and the ``No-hair" theorem. Despite several observational efforts to detect such systems, none have been found so far. One explanation for this maybe the extreme Doppler smearing caused as the pulsar moves around its orbit during the observation time. The classical approach to account for this is do an acceleration search \citep{1991ApJ...368..504J} i.e to first order assume that the acceleration of the binary pulsar is constant during the observation. This assumption holds weight only if the observation samples a small fraction of the orbit typically $P\mathrm{_{orb}}~\geq$~10~$T\mathrm{_{obs}}$ \citep{2003ApJ...589..911R, 2015MNRAS.450.2922N}. Therefore, in order to be sensitive to highly relativistic systems we would require very short observations. This however increases the minimum flux density of the pulsar that we can detect as sensitivity grows with $T\mathrm{_{obs}}^{1/2}$. Therefore, by using an acceleration search we have a trade-off between sensitivity and the ability to detect short-orbital period binaries. Another option is to do a higher-order polynomial search, for example a jerk search which assumes a constant jerk during the observation. \citet{2018ApJ...863L..13A} showed that using this technique leads to an increased sensitivity to short orbital period binaries $P\mathrm{_{orb}} \approx$ 5-15 $T\mathrm{_{obs}}$. \citet{2013MNRAS.431..292E} also demonstrated the improvements of jerk searches compared to acceleration searches for observations of \mbox{PSR~J0737-3039A}. \citet{2016arXiv161006831S} proposed four new algorithms for radio pulsar searches including a semicoherent search which divides the data into small chunks and combines information from coherent subsearches while preserving as much phase information as possible. Yet another approach is do a fully coherent search on the full length observation by searching directly over the Keplerian parameters. Assuming a circular orbit binary, this would lead to a three-parameter search over orbital period, projected semi-major axis and initial orbital phase. \citet{2011PhDT.......293K, 2013ApJ...773...91A} and \citet{2013ApJ...774...93K} did exactly this by using a matched filtering process, convolving the data with circular orbit templates in order to find pulsars in Pulsar Arecibo L-band Feed Array (PALFA; \citealt{2006ApJ...637..446C}) and Parkes Multibeam Pulsar Survey (PMPS; \citealt{2001MNRAS.328...17M}) survey using the volunteer-distributed Einstein@Home project. Assuming standard binary stellar evolution model \citep{1998astro.ph..1127Y, 2003MNRAS.342.1169V}, PSR-BH binaries are expected to be in highly eccentric orbits \citep{2018MNRAS.477L.128S}. While, we are already computationally limited with the acceleration and jerk search techniques, a yet unexplored question is how feasible would a fully coherent elliptical orbit search (five-parameter search) be and if there are any advantages in terms of sensitivity  towards unexplored parameter spaces that a Keplerian search would provide. Recent advances in graphics processing unit (GPU) technologies have considerably sped up the typically computationally expensive parts of pulsar search pipelines like dedisperison, time/frequency domain resampling, calculating a Fast-Fourier transform (FFT) and harmonic summing. Here, we present a GPU-based implementation of a binary pulsar search pipeline that searches across three and five Keplerian parameters using the template-bank algorithm and we demonstrate that it is more sensitive than acceleration and jerk search pipelines to PSR-SBH binaries in eccentric orbits. We do a detailed comparison of the detectability of mildly eccentric PSR-SBH binaries using our pipeline and compare to all of the currently used standard binary pulsar pipelines including those mentioned above. We also give rough scaling relations of the dependence of total orbital templates on the Keplerian search parameters. Finally, we quantify the computational feasibility of using this method for targeted and wide-area galactic plane observations in the near future. 

\section{Template-Bank Search}
\subsection{Signal Model and Detection Statistic}
To discover new pulsars buried in our observations, we need to first define a signal model. We make the following assumptions in our signal model. The orbit of the pulsar is Keplerian.\footnote{This assumption is based on the typical timescales of search-mode observation of radio pulsars which is in the order of few hours. For long observations lasting months or years, additional parameters including Post-Keplerian and spin-down parameters need to be added to the phase model.}  We know the exact location of the source on the sky, therefore Doppler phase drifts caused due to the detector motion can be removed and the signal is monochromatic at the spin-frequency of the pulsar $f_0$ i.e we ignore spin-down effects of the pulsar. Therefore, the rotational phase $\Phi$ for the fundamental mode of the signal emitted by a uniformly rotating pulsar in an eccentric orbit observed from solar-system barycenter is
\begin{equation}\label{eq:main_rotational_phase}
\Phi(t; \Lambda) = 2 \pi f_0 t + \phi_D (t),
\end{equation}
where t is the barycentric time, $f_0$ is the spin-frequency of the pulsar, $\Lambda$ is a tuple of the unknown binary parameters of the pulsar signal and $\phi_D (t)$ is the Doppler phase correction factor due to the pulsar being in an eccentric orbit, given by (see section 2B of \citet{2001PhRvD..63l2001D} for the full derivation), 
\begin{equation}\label{eq:doppler_elliptical}
    \phi_D(t; \Lambda) = \frac{2\pi f_0 a \sin \epsilon}{c} \big[\sin \psi \cos E(t) + \cos \psi \sqrt{1 - e^2} \sin E(t)\big].
\end{equation}

Here, $E(t)$ is the eccentric anomaly which is a function of time, $a$ is the semi-major axis of the orbit, $\epsilon$ is the inclination angle and $\frac{asin{\epsilon}}{c}$ is the projected semi-major axis of the orbit in light-seconds. $\psi$ is the longitude of periastron and $e$ is the eccentricity of the orbit. The eccentric anomaly $E(t)$ is related to the mean angular velocity $\Omega$ and the mean anomaly $M(t)$ by the Kepler equation,
\begin{equation}
E(t) - e \sin E(t) = \Omega t + \alpha \approx M(t),
\end{equation}
where $\alpha$ is the initial orbital phase of the pulsar orbit. We have seven unknown parameters $\Lambda = \{f_0, a, \epsilon, \Omega, \alpha, \psi, e\}$ in equation \ref{eq:main_rotational_phase}. The semi-major axis of the pulsar cannot be directly observed. Since we search over different waveforms in our signal model, what matters for our search is the combination of parameters that affect the rotational phase, in this case the projected semi-major axis $\tau~=~\frac{a \sin \epsilon}{c}$. Therefore, in practice an untargeted coherent search for binary pulsars in a Keplerian orbit needs to search over the six-dimensional parameter space of $\Lambda = \{f_0, \tau, \Omega, \alpha, \psi, e\}$. It is unlikely that without any prior information, we would exactly sample the true parameters of the undiscovered pulsar, therefore we use a matched filtering process of convolving the data with multiple waveforms. Each sample trial consists of different six-tuple parameter combinations and is called a template of the signal. A combination of many such templates form a template bank. The density of templates in the template-bank can be characterised by its mismatch value (defined in equation \ref{eq:mismatch}) i.e the worst-case detection statistic loss. Our general goal is to cover large parts of the parameter space with the minimum number of templates possible in order to minimise computational costs. This is a widely used technique especially in gravitational-wave analysis, see \citet{1999PhRvD..60b2002O,2007PhRvD..76h2001A,2009PhRvD..80d2003A}. In order to sample in this parameter space efficiently, we need to define distances between two points. This is given by the parameter space metric \citep{1996PhRvD..53.6749O} defined in the next section. For completeness we also describe the signal phase assuming a circular orbit. In this case eccentricity is zero and the longitude of periastron and the eccentric anomaly are undefined, therefore the phase model becomes
\begin{equation}
\phi(t; \Lambda) = 2 \pi f_0  \big[t + \tau \sin(\Omega t + \alpha)\big].
\label{eq: circular phase model}
\end{equation}
In this case we search over the four-dimensional parameter space of $\Lambda = \{f_0, \tau, \Omega, \alpha\}$.

The total time-domain radio intensity signal is a sum of instrumental and environmental noise $N(t)$ and harmonic sums of a pulsar signal.
\begin{equation}
s(t;\Lambda) \equiv N(t) + \sum_{n = 1}^{\infty} s_n (t; \Lambda),
\end{equation}
where the intensity of each harmonic is given by 
\begin{equation}
s_n(t; \Lambda) \equiv \Re\big[\mathcal{A}_n \mathrm{exp} [-in \Phi(t;\Lambda)]\big],
\end{equation}
where $\mathcal{A}_n$ are the complex amplitudes of the harmonics of the signal. We then define a coherent detection statistic $P_n (\Lambda, \Lambda^{\prime})$ for the nth harmonic computed using the radio intensity correlated with the nth normalised signal template currently being searched exp$[-in \Phi(t; \Lambda)]$. This is the detection statistic\footnote{Throughout this article T and T$\mathrm{_{obs}}$ have been used interchangeably to refer to the observation time of a pulsar-search mode observation.} recovered from a pulsar signal with true parameters at $\Lambda$ with a template at $\Lambda^{\prime}$. 
\begin{equation}
P_n (\Lambda, \Lambda^{\prime}) = \left |{\frac{1}{T} \int_{0}^{T} dt \, s(t;\Lambda) \, \mathrm{exp} \left[-in \Phi (t; \Lambda^{\prime}) \right]  }\right |^2.
\end{equation}
Thresholding the detection statistic helps us minimize the false-alarm probability at a fixed rate of false-alarm probability \citep{2002PhRvD..66j2003A}. In the absence of a pulsar signal, the $N(t)$ term dominates and the detection statistic term is proportional to the instrumental noise whereas in the presence of a strong signal the $N(t)$ term can be neglected and the expectation value of the detection statistic then becomes,

\begin{equation}
\label{eq: power detection statistic}
\langle P_n(\Lambda, \Lambda^{\prime}) \rangle \approx \left | \frac{A_n}{2} \right |^2 \left | \frac{1}{T} \int_0^T dt \, \exp [in(\Phi(t; \Lambda) - \Phi(t; \Lambda^{\prime}))] \right |^2.
\end{equation}
 While searching for new pulsars, since we do not known a \emph{priori} the pulse-profile, we assume the Fourier transform of the pulse profile resembles a Dirac delta function truncated to a finite number of harmonics. Therefore, we equally weight $P_n$ for different harmonics (P$_0$,..P$_4$) and sum them together to form our combined detection statistic.

\begin{equation}
    S_L = \sum_{n = 1}^{2^L}P_n.
\end{equation}
where 2$^L$ is the number of harmonic sums performed. In the noise-only case, assuming Gaussian statistics, the probability distribution function (PDF) of $S_L$ is a $\chi^2$ distribution with 2N = 2$^{L+1}$ degrees of freedom. Integrating this PDF shows that the probability for the power in any spectral bin to exceed a threshold $P_{\rm min}$ is proportional to exp(-$P_{\rm min}$). This is called the false-alarm probability i.e the chance that a candidate arises from random noise fluctuations rather than a true pulsar signal. 
$P_{\rm min}$ can be estimated by setting the number of false postive to one.
\begin{equation}
P_{\rm min} = -\ln\left({1/(2 n_{\rm samp})}\right),
\end{equation}
where $n_{\rm samp}$ is the number of samples in the timeseries. \citet{2012hpa..book.....L} showed that we can convert the power threshold $P_{\rm min}$ into a Signal to Noise ratio threshold S/N$_{\rm min}$ given by
\begin{equation}
 \rm{S/N}_{\rm min} = \frac{\sqrt{\ln{\left[n_{\rm trials}\right]}} - \sqrt{\pi/4}}{\sqrt{1 - \pi/4}},
\end{equation}
where $n_{\rm trials} = n_{\rm samp} \times n_{\rm DM \, trials} \times n_{\rm harmonic\, sums} \times n_{\rm orbital\, trials}$. In practise, due to the presence of Radio frequency interference (RFI) a higher threshold maybe required. We consider all candidates with Signal to Noise ratio greater than S/N$_{\rm min}$ to be statistically significant and visually inspect them. 
\subsection{Definition of Metric and Mismatch}
\label{sec: metric}
Using equation \ref{eq: power detection statistic}, we can define the mismatch (fractional loss of detection statistic, in our case signal to noise ratio) between two points as 
\begin{equation}
\label{eq:mismatch}
m(\Lambda, \Lambda^{\prime}) = 1 - \frac{P_n (\Lambda, \Lambda^{\prime})}{P_n (\Lambda, \Lambda)} \approx g_{\alpha \beta} \, \Delta \Lambda^{\alpha} \Delta   \Lambda^{\beta} + \mathcal{O}(\Delta \Lambda^3),
\end{equation}
where a mismatch of one implies complete loss of signal and a mismatch of zero implies a perfect recovery of the signal, $g_{\alpha \beta}$ is the metric tensor, $\alpha$ and $\beta$ correspond to points in the six-parameter space described earlier and we adopt the Einstein summation convention where repeated indices are summed over. For small deviations of the parameter space coordinates, $g_{\alpha \beta}$ can be calculated by Taylor expansion as (section 8.5 of \citet{2011PhDT.......293K} for the full derivation)
\begin{equation}
\label{eq:metric tensor}
g_{\alpha \beta} = \langle \partial_{\alpha } \Phi \partial_{\beta} \Phi  \rangle_T - \langle \partial_{\alpha} \Phi \rangle_T \langle \partial_{\beta} \Phi \rangle_T,
\end{equation}
where the angle brackets represent time average of a function $G(t)$ 
\begin{equation}
\langle G(t) \rangle_T \equiv \frac{1}{T} \int_0^T  dt G(t).
\end{equation}
For computing the time-averages of the derivatives of the phase with respect to the search parameters, it is convenient to express $\cos E$ and $\sin E$ as a power series in the eccentricity parameter $e$ with harmonics in the mean anomaly $M$
\begin{equation}
\cos E = \sum_{k=0}^{\infty} C_k (e) \cos(kM),
\label{eq: cosE ell}
\end{equation}

\begin{equation}
\sqrt{1 - e^2}\sin E = \sum_{k=1}^{\infty} S_k (e) \sin(kM),
\label{eq: sinE ell}
\end{equation}
where C$_k$ and S$_k$ are power series in $e$. The order to which one should consider expanding the power series depends on the spin-frequency and eccentricity we are interested in searching for. See Appendix C of \citet{2020ApJ...901..156N} for a derivation of this power series. In this work we consider up-to the 7th power in $e$, and these values can be found in Appendix \ref{sec: coefficients}.

\subsection{Random Template Banks}
The central idea behind this algorithm is to give up the requirement of complete coverage of the parameter space ($\eta$ = 1). Instead, we aim for a user-defined probabilistic coverage of the parameter space ($\eta$~<~1). The templates are distributed randomly with uniform probability based on values of the square root of the determinant of the parameter space metric. We refer the reader to \citet{2009PhRvD..79j4017M} for an in-depth review of this algorithm including derivations for the expressions mentioned here. Here we summarise the relevant details that are applicable to our work. Using the definition of mismatch (described in equation \ref{eq:mismatch}), we can define a metric for our parameter space $g_{\alpha \beta}$ for different signal parameter values $(\lambda_1, \lambda_2, ..., \lambda_n)$ and calculate the proper volume $V_{S_n}$ of our parameter space $S_n$ by:

\begin{equation}
V_{S_n} = \int_{S_n} \, dV \, = \int_{S_n} \, d^n \lambda \sqrt{g},
\label{eq:volume}
\end{equation}
where $g = \det(g_{\alpha \beta}$) is the determinant of the parameter space metric. We then define the volume of an n-dimensional ball with unit radius as:
\begin{equation}
C_n = \frac{\pi^{\frac{n}{2}}}{\Gamma \left (\frac{n}{2} + 1 \right)},
\end{equation}
This is the n-dimensional volume enclosed by a (n-1) dimensional sphere. The volume covered by a single template with mismatch $m_{\ast}$ is:
\begin{equation}
V_T = C_n m_*^{\frac{n}{2}}.
\end{equation}
The total number of random templates required to achieve a coverage $\eta$ with nominal mismatch $m_{\ast}$ of a parameter space $S_n$ is thus:
\begin{equation}
\label{eq:total templates}
N_R (\eta, m_{\ast}, S_n) = \frac{\ln(1 - \eta)}{\ln \left (1 - m_{\ast}^{\frac{n}{2}} \frac{C_n}{V_{S_n}} \right)}.
\end{equation}
In most cases, we are interested in large parameter spaces where the volume of the parameter space $V_{S_n}$ is much greater than the volume of a single template $V_T$, therefore, $m_{\ast}^{\frac{n}{2}} \frac{
C_n}{V_{S_n}} \ll 1$, and we can Taylor-expand equation \ref{eq:total templates} to write,
\begin{equation}
N_R (\eta, m_{\ast}, S_n) \approx \frac{m_{\ast}^{\frac{-n}{2}} V_{S_n}}{C_n} \ln \left(\frac{1}{1 - \eta} \right ).
\label{eq:number templates}
\end{equation}
\subsection{Frequency-Projected Metric}
Our template bank construction method is similar to previous works by \citet{2011PhDT.......293K, 2013ApJ...773...91A} and \citet{2013ApJ...774...93K}. We do this by computing a Cartesian product of a five-dimensional orbital template bank $\Lambda_{\rm orb} = \{\tau, \Omega, \alpha, \psi, e\}$ along with a uniformly spaced grid in the spin-frequency axis. This is mainly done for computational reasons as the latter can be efficiently calculated using the FFT algorithm \citep{CooleyTukey}. Therefore, we take the metric tensor (defined in equation \ref{eq:metric tensor}) and project it to a subspace of constant spin frequency $f_0$. This can be thought of as a slice across the parameter-space for a given $f_0$. This is usually chosen to be the highest spin-frequency of the harmonic of the signal we would like to detect in our search $f_{\rm max}$. This number should be chosen carefully as the number of templates for a five-dimensional coherent elliptical orbit search grows to the fifth power of spin frequency $\left(\mathrm{N_{templates} \propto f_0^5} \right)$. Choosing a particular $f_{\rm max}$ does not automatically imply that we cannot detect signals with $f_0 > f_{\rm max}$. This usually means that our mismatch values would be higher for higher spin frequencies and conversely our mismatch values will be lower for lower spin frequencies. In other words, this means that we over-sample our parameter space for lower spin frequency values, under-sample for higher spin-frequency values and we are optimally sampled for the $f_{\rm max}$ we choose to create our template-bank. We calculate the frequency projected metric as

\begin{equation}
 \gamma_{\alpha \beta} = g_{\alpha\beta} - \frac{g_{f\!\alpha} g_{f\!\beta}}{g_{f\!f}},
\end{equation}
where the repeated indices are summed over. Therefore, for each orbital template, we apply a time-domain resampling algorithm (described in Appendix \ref{sec:software}) and then calculate an FFT.

\begin{table*}
\begin{center}
\caption{Parameter values and prior ranges used for generating our orbital template bank. U here refers to a uniform probability distribution.}
\label{tab:prior ranges}
\begin{tabular}{ccc}
\hline
Parameter & Value/Range & Unit \\

\hline
$t\mathrm{_{obs}}$ & 1.2 & h  \\
$\mathrm{m_{pulsar, min}}$ & 1.4 & $M_{\odot}$ \\
$\mathrm{m_{companion, max}}$ & 8.0 & $M_{\odot}$ \\
Coverage ($\eta$) & 0.9 & \\
Mismatch (m) & 0.2 \\
Max.Spin Frequency (f$\mathrm{_{max}}$) & 66.67 & Hz \\
Orbital Period ($\mathrm{P_{orb}}$) & U(6, 12) & h \\
Projected Radius ($\tau$) & \begin{minipage}{4cm}{\begin{equation*}
\mathrm{U}\left({0, \frac{G^{\frac{1}{3}} \Omega_{\rm orb}^{-\frac{2}{3}} m_{\rm comp,max}}{c \left(m_{\rm pulsar, min} + m_{\rm comp,max}\right)^{\frac{2}{3}}}}\right)
\end{equation*} }\end{minipage}
& lt-s \\
Initial Orbital Phase  ($\alpha$) & U(0, 2 $\pi$) & rad \\
Longitude of Periastron ($\psi$) & U(0, 2 $\pi$) & rad \\
Eccentricity (e) & U(0, 0.1) & \\
 
\hline

\end{tabular}
\end{center}
\end{table*}

\subsection{Signal Phase and Dimensionless Parameters}
Following the formalism described in \citet{2001PhRvD..63l2001D}, we re-write our signal phase in dimensionless parameters in order to simplify the calculation for the determinant of the parameter space metric. If we multiply our signal phase model $\Phi$ by a constant factor $\chi$, such that
\begin{equation}
\Phi = \chi \tilde{\Phi},
\end{equation}
then each component of the metric $g_{\alpha \beta}$ and the frequency-projected metric $\gamma_{\alpha \beta}$ become
\begin{equation}
g_{\alpha \beta} = \chi^2 \tilde{g}_{\alpha \beta}, \quad \gamma_{\alpha \beta} = \chi^2 \tilde{\gamma}_{\alpha \beta},
\end{equation}
and the determinants are scaled by 
\begin{equation}
det(g_{\alpha \beta}) = \chi^{2N + 2} \det(\tilde{g}_{\alpha \beta}), \quad det(\gamma_{\alpha \beta}) = \chi^{2N} \det(\tilde{\gamma}_{\alpha \beta})
\end{equation}
As shown in equation \ref{eq:volume} and \ref{eq:number templates}, the proper volume and the total number of templates is proportional to the square root of the determinant of the metric tensor, which in these coordinates becomes,
\begin{equation}
V_{S_n} = \chi^N \tilde{V}_{S_n}, \quad N_R = \chi^N \tilde{N}_R.
\label{eq: final templates equation}
\end{equation}
where the `N' in the exponent refers to the number of free parameters in our signal model. In the circular case N=3 and in the elliptical case N=5. Using these properties, we re-write our signal phase model as
\begin{equation}
\Phi = (2 \pi f_0 T) \tilde{\Phi},
\label{eq: dimensionless scaling factor}
\end{equation}
where 
\begin{equation}
\tilde{\Phi} = \kappa u + X \cos E + Y \sqrt{1 - e^2} \sin E.
\label{eq: dimensionless phase}
\end{equation}
The dimensionless parameters are defined as,
\begin{equation}
\begin{aligned}
u &= \frac{t}{T},\\
\kappa &= \frac{f - f_0}{f_0},\\
X &= \frac{\tau \sin \psi}{T}, \\
Y &= \frac{\tau \cos \psi}{T}, \\
\Omega^{\prime} &= \Omega T.
\end{aligned}
\end{equation}
Here $u$ is dimensionless time, satisfying $0 \leq u \leq 1$, $\kappa$ is dimensionless frequency satisfying  $-1\leq\kappa\leq0$. $\Omega^{\prime}$ is the number of orbits in radians and $\frac{\Omega^{\prime}}{2 \pi}$ is the number of orbits covered by the binary during the observation time T From equation \ref{eq: final templates equation} and \ref{eq: dimensionless scaling factor}, it is clear that
\begin{equation}
\chi = 2 \pi f_{\rm max} T,
\end{equation}
where we substitute $f_{0} = f_{\rm max}$ which is the maximum spin-frequency of the pulsar we want our search to be sensitive to with the chosen mismatch. Using the dimensionless rotational phase (defined in equation \ref{eq: dimensionless phase}) we then proceed to calculate the metric tensor $g_{\alpha \beta}$ (using equation \ref{eq:metric tensor}) with parameters $\Lambda = \{\kappa, X, Y, e, \Omega, \alpha\}$ and frequency-projected metric $\gamma_{\alpha \beta}$ with parameters $\Lambda = \{X, Y, e, \Omega, \alpha\}$. We then calculate the square root of the determinant of this frequency projected metric tensor. Finally, we multiply this value by our scaling factor $(2 \pi f_{\rm max} T)^5$. 

\begin{figure*}
	\includegraphics[width=2.0\columnwidth]{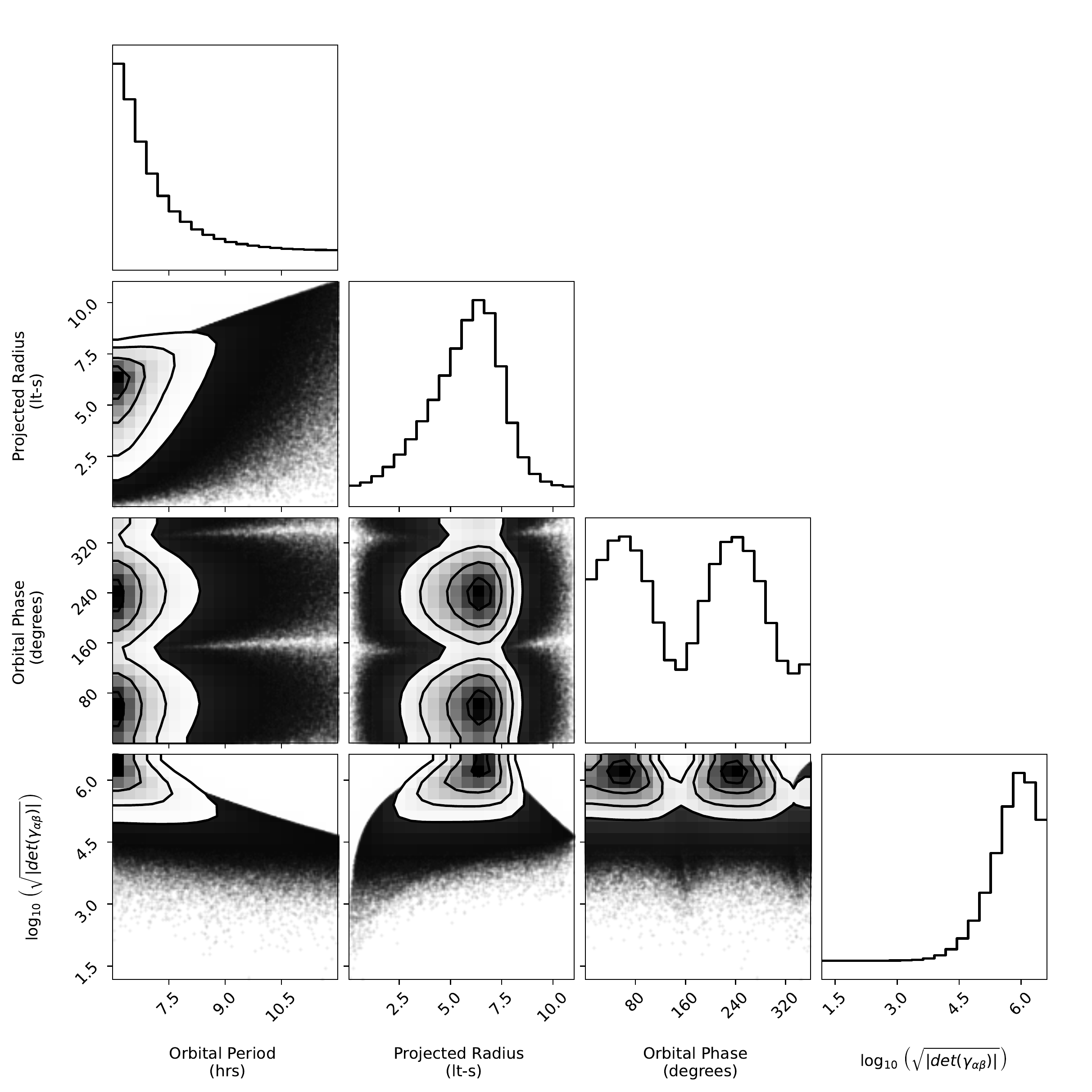}
    \caption{Corner plot showing the distribution of the value of  $\sqrt{|\gamma_{\alpha \beta} |}$ which is a measure of our template pseudo-density for different orbital parameter combinations using the circular orbit phase model. Black points show the relative sampling density in the template bank. Histograms in the main diagonal show the one-dimensional marginalized distribution of the template density across each of our orbital parameters. We see that there is exponential growth of orbital templates for short orbital periods and high $a \sin \epsilon$ values. Contours in the 2-D plots highlight values lying with 0.5, 1, 1.5 and 2 sigma respectively. $a \sin \epsilon = 0$ is the special case of face-on orbit when there is no Doppler modulation of the signal. In the orbital phase versus orbital period plot, we see two streaks of low values of $\sqrt{|\gamma_{\alpha \beta} |}$. This corresponds to $\left (1 - \frac{T\mathrm{_{obs}}}{P\mathrm{_{orb}}} \right) \times 180$ and $\left (2 - \frac{T\mathrm{_{obs}}}{P\mathrm{_{orb}}} \right) \times 180$ degrees. At these orbital phases, the pulsar is either approaching towards or receding from the observer essentially along the line of sight, and the determinant tends to zero. This makes the signal appear Doppler shifted by a constant, depending on the velocity of the source that does not need to be corrected for.}
    \label{fig:corner plot circular}
\end{figure*}

\begin{figure*}
	\includegraphics[width=2.2\columnwidth]{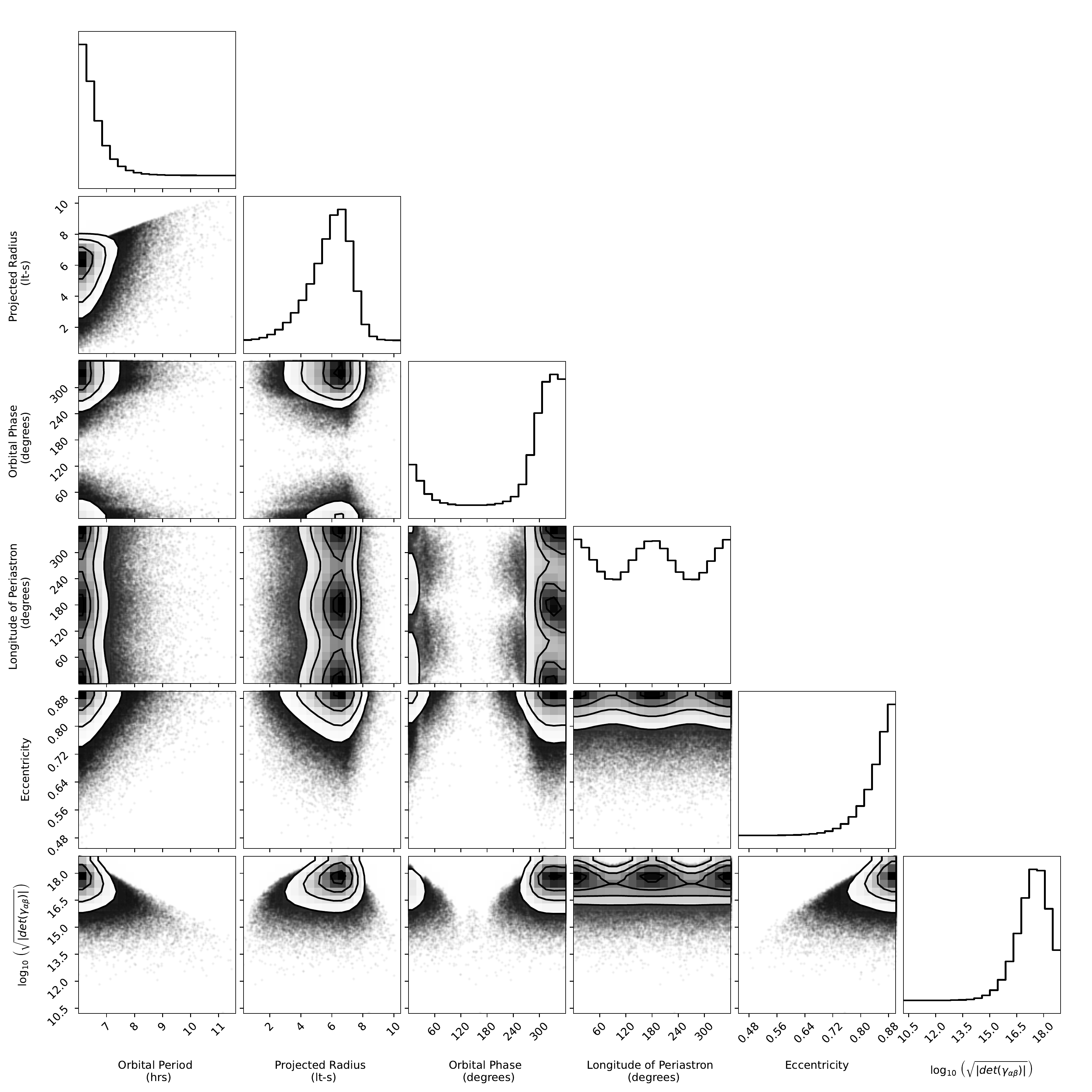}
    \caption{Corner plot showing the distribution of the value of  $\sqrt{|\gamma_{\alpha \beta} |}$ which is a measure of our template pseudo-density for different orbital parameter combinations using the elliptical orbit phase model with maximum eccentricity e$_{\rm max} = 0.9$. Black points show the relative sampling density in the template bank. As in the circular case, there is exponential increase in the number of orbital templates for a combination of short orbital periods and high $a \sin \epsilon$ values. There is also an exponential increase in templates across the eccentricity parameter and a roughly cosine dependence across the longitude of periastron parameter. In terms of orbital phase, most of the templates are distributed around a region covering 180$^{\circ}$ degrees.  Contours in the 2-D plots highlight higher density regions of orbital templates. These contours highlight values lying within 0.5, 1, 1.5 and 2 sigma respectively.}
    \label{fig:corner plot ellptical}
\end{figure*}

\begin{table*}
\begin{center}
\caption{Input, Spin and Orbital Parameters for the Simulated Pulsar-Black hole binaries used in our analysis. We simulated all our orbits with a Dispersion Measure DM = 0. Here, `U' denotes a uniform distribution.}
\label{tab:simulation parameters}
\begin{tabular}{ccc}
\hline
Parameter & Value/Range & Unit \\

\hline
$t\mathrm{_{obs}}$ & 1.2 & h  \\
$\mathrm{m_{pulsar, min}}$ & 1.4 & $M_{\odot}$ \\
$\mathrm{m_{companion, max}}$ & 8.0 & $M_{\odot}$ \\
Spin-Period  & 15.0 & ms \\
Duty Cycle  & 10.0 & \% \\
Dispersion Measure  & 0.0 & \SI{} {\parsec \per \centi \meter \cubed} \\
Frequency Channels  & 1024 &  \\
Bandwidth  & 400 & MHz \\
Frequency of Channel1  & 1181.804688 & MHz \\
Channel bandwidth & 0.390625 & MHz \\
Signal-to-noise ratio of single pulse & 0.02 & \\
Data Bit-Rate & 4 \\
Number of samples & 2$^{24}$ & samples \\ 
Sampling time & 256e-06 & s\\ 
Orbital Period ($\mathrm{P_{orb}}$) & U(6, 12) & h \\
Eccentricity (e) & 0.1 & \\
Initial Orbital Phase  ($\alpha$) & U(0, 2 $\pi$) & rad \\
Longitude of Periastron ($\psi$) & U(0, 2 $\pi$) & rad \\
Projected Radius ($\tau$) & \begin{minipage}{4cm}{\begin{equation*}
\mathrm{U}\left({0, \frac{G^{\frac{1}{3}} \Omega_{\rm orb}^{-\frac{2}{3}} m_{\rm comp,max}}{c \left(m_{\rm pulsar, min} + m_{\rm comp,max}\right)^{\frac{2}{3}}}}\right)
\end{equation*} }\end{minipage}
& lt-s \\
 
\hline

\end{tabular}
\end{center}
\end{table*}

\subsection{Chosen Parameter-Space}
Before we start constructing a template bank to search for binary pulsars, we first need to decide what region of the parameter space will we focus our search on.  The two main factors that are important for this are our astrophysical motivation and computational feasibility. The former is based on our prior information about binary stellar evolution and past searches on our data which can help us exclude regions that are unlikely to yield promising results or already been investigated. The most important criteria is how much computational resource is available and how much time can be invested in doing our search analysis. With unlimited computing power we could explore all regions of our parameter space, however in practice we try to balance both these factors by maximising our chances of detecting an exciting binary pulsar at a fixed computational cost. 

The most commonly used search techniques to find binary pulsars in radio observations are acceleration searches which can done in both the time domain \citep{1991ApJ...368..504J} and frequency domain \citep{2002AJ....124.1788R}. Recently, \textsc{PRESTO}\footnote{\url{https://github.com/scottransom/presto}} has added support for frequency-domain jerk searches \citep{2018ApJ...863L..13A}. \citet{2013ApJ...773...91A, 2013ApJ...774...93K} did a coherent search using the template-bank algorithm described earlier (assuming circular orbits - three orbital search parameters)  to search for compact binaries within PALFA \citep{2006ApJ...637..446C} and PMPS \citep{2001MNRAS.328...17M} observations respectively. Given the extensive amount of work that has already been done building and testing these software suites and their remarkable success in finding several relativistic binaries like PSR~J0737-3039A \citep{2003Natur.426..531B} which was discovered in a FFT search, PSR~J1757-1854 \citep{2018MNRAS.475L..57C} which was discovered in a segmented acceleration search among others, our goal here is to complement their work by focusing our search on regions of parameter space that were not searchable by these algorithms. 

In Section \ref{sec:compare} we do a detailed simulation comparing these search algorithms. We fix our upper limit for orbital period as $P_{\rm orb} = 10 T$. This is because the assumption of ``constant'' acceleration works well for long orbital period binaries i.e when $P_{\rm orb} > 10 T$. The lower limit of the orbital period searched is determined solely by the amount of computational power available. The constraint on projected semi-major axis, $a \sin \epsilon$, is calculated based on the orbital period, masses of the pulsar and companion, and the different inclination angles we want to be sensitive to. Using Kepler's third law, we define
\begin{equation}
0 \leq a \sin{\epsilon} \leq \theta \frac{G^{\frac{1}{3}} \Omega_{\rm orb}^{-\frac{2}{3}} m_{\rm comp,max}}{c \left(m_{\rm pulsar, min} + m_{\rm comp,max}\right)^{\frac{2}{3}}} \mathrm{lt\mbox{-}s}
\end{equation}

where $m_{\rm comp,max}$ is the maximum companion mass, $m_{\rm pulsar, min}$ is the minimum pulsar mass, $G$ is the gravitational constant, $c$ is the speed of light in vacuum, the parameter $\theta$ ($0 \leq \theta \leq 1$) constraints the inclination angle that we want to be sensitive towards for a given binary system. In our template bank, we selected the following binary parameter ranges: $m_{\rm pulsar, min}=1.4 \, M_{\odot}, m_{\rm comp,max}=8.0 \, M_{\odot}, \, \theta=1, P_{\rm orb} =$ 6-12 hrs. We had no constraints on the initial orbital phase $\alpha$ ( $0 \leq \alpha \leq 2 \pi$) and longitude of periastron $\psi$ ($0 \leq \psi \leq 2 \pi$). The upper-limit on the eccentricity range is also decided by computational factors. We limit our searches to a range between $0 \leq e \leq 0.1$. The prior ranges used in our tests are summarised in table \ref{tab:prior ranges}.
\subsection{Distributing templates in the parameter space}
Once we have calculated the number of required orbital templates using equations \ref{eq:volume} and \ref{eq:number templates}, the next step is to generate the template-bank, i.e.\ distribute the orbital templates in the parameter space. Our goal is to sample from a probability density function that is proportional to the square root of the determinant of the metric tensor $\sqrt{|\gamma_{\alpha \beta} |}$ multiplied by the prior probabilities in our orbital parameter range. Markov Chain Monte Carlo (MCMC) methods are well suited for such problems. We use the \textsc{Python} package \textsc{emcee} \citep{emcee} which is an implementation of the affine-invariant MCMC ensemble sampler proposed by \citep{2010CAMCS...5...65G}. Briefly, this method works as follows. We explore the parameter space through a set of ``walkers'' which run in parallel. Each walker represents a point in the parameter space. After each iteration, a walker selects another walker and takes a step along the line in parameter space connecting between them. These step sizes are chosen stochastically. At this new step, we then calculate our posterior distribution function (PDF). We always accept steps where the PDF value increases and sometimes accept steps if the PDF value decreases. These walkers can be thought of as separate Metropolis-Hastings chains running in parallel with the caveat that the steps taken by one walker are dependent on the position of the other walkers. Based on the recommendation in \citet{2010CAMCS...5...65G}, we use the integrated autocorrleation time to quantify the effects of sampling error in our chain. We assume our chains have converged when the length of the chain is a fifty times our integrated autocorrleation time averaged across all dimensions. We chose 800 walkers for our chain, and once the chains have sufficiently converged, we discard the burn-in samples and randomly select the number of required templates from our chain. Using this method, we generated two orbital template banks, one assuming a circular orbit and another assuming an elliptical orbit.

Figure \ref{fig:corner plot circular} shows a corner plot \citep{corner} for the circular orbit case. Plots along the longest diagonal show the one dimensional marginalized distribution of the template pseudo density $\sqrt{|\gamma_{\alpha \beta} |}$ in base 10 log units for each of the orbital parameters. These plots give us an indication as to how the templates are distributed across the parameter space.\footnote{Templates with higher values of $\sqrt{|\gamma_{\alpha \beta} |}$ are more likely to be accepted.} The off-diagonal plots show the marginalized two dimensional distributions for different combinations of the orbital parameters. We see that the number of accepted/required templates grows exponentially for shorter orbital periods and a combination of short orbital periods and high-a$\sin \epsilon$ values. We see a dip in the template density for high values of a$\sin \epsilon$ because longer orbital period binaries tend to higher a$\sin \epsilon$ value ranges due to Kepler's third law. In the orbital phase versus orbital period plot, we see two streaks of very low values of $\sqrt{|\gamma_{\alpha \beta} |}$. This corresponds to $\left (1 - \frac{T\mathrm{_{obs}}}{P\mathrm{_{orb}}} \right) \times 180$ and $\left (2 - \frac{T\mathrm{_{obs}}}{P\mathrm{_{orb}}} \right) \times 180$ degrees. At these orbital phases, the pulsar is either approaching or receding from the observer essentially along the line of sight, and the determinant tends to zero. This makes the signal appear Doppler shifted by a constant, depending on the velocity of the source that does not need to be corrected for.

Figure \ref{fig:corner plot ellptical} shows the corner plot for the elliptical orbit case. Here, we show the scaling of $\sqrt{|\gamma_{\alpha \beta} |}$ up-to a maximum eccentricity of e$_{\rm max} = 0.9$ for the same orbital period range of 6-12 hours. Here, we see that the maximum value of $\sqrt{|\gamma_{\alpha \beta} |}$ has gone up by twelve orders of magnitude which gives us an indication of the extremely high number of orbital templates that would be required to conduct such a search. We will expand more on scaling relations and computational feasibility of these searches in sections \ref{sec:scaling relation keplerian} and \ref{sec: suitability}. Similar to the circular orbit case, the plots along the diagonal show the one-dimensional marginalized distribution of $\sqrt{|\gamma_{\alpha \beta} |}$ for each of the five orbital parameters. We see a similar increase in the number of trials for short orbital periods and a combination of short orbital periods and high a$\sin \epsilon$ values. There is also an exponential dependence of the template density across the eccentricity parameter. Most of the templates are distributed around a region covering 180$^{\circ}$ across initial orbital phase. Additionally, we also notice a near cosine dependence of template density across the longitude of periastron parameter. For the elliptical orbit searches used in this paper, we fix the e$_{\rm max} = 0.1$ in order to do a computationally feasible search which leads to an increase of slightly more than 2 orders of magnitude more trials than than the circular orbit case.

Using the metric approximation defined in section \ref{sec: metric} leads to an over-estimation of the required orbital templates in the template bank. One method to reduce the total number of orbital templates is to apply the Stochastic template bank algorithm \citep{2009PhRvD..80j4014H}. Stochastic template banks are formed in a similar manner with an additional second step that prunes orbital templates that are closer than the nominal mismatch value. We provide an implementation of this algorithm in our software suite. Creating a stochastic template bank is computationally expensive. Therefore, the trade-off between calculating a stochastic template bank and time-saved from search analysis should be evaluated. For all our tests in the subsequent sections we use a random template bank to save compute time.

\subsection{Template Bank Verification}
\begin{figure}
	\includegraphics[width=1\columnwidth]{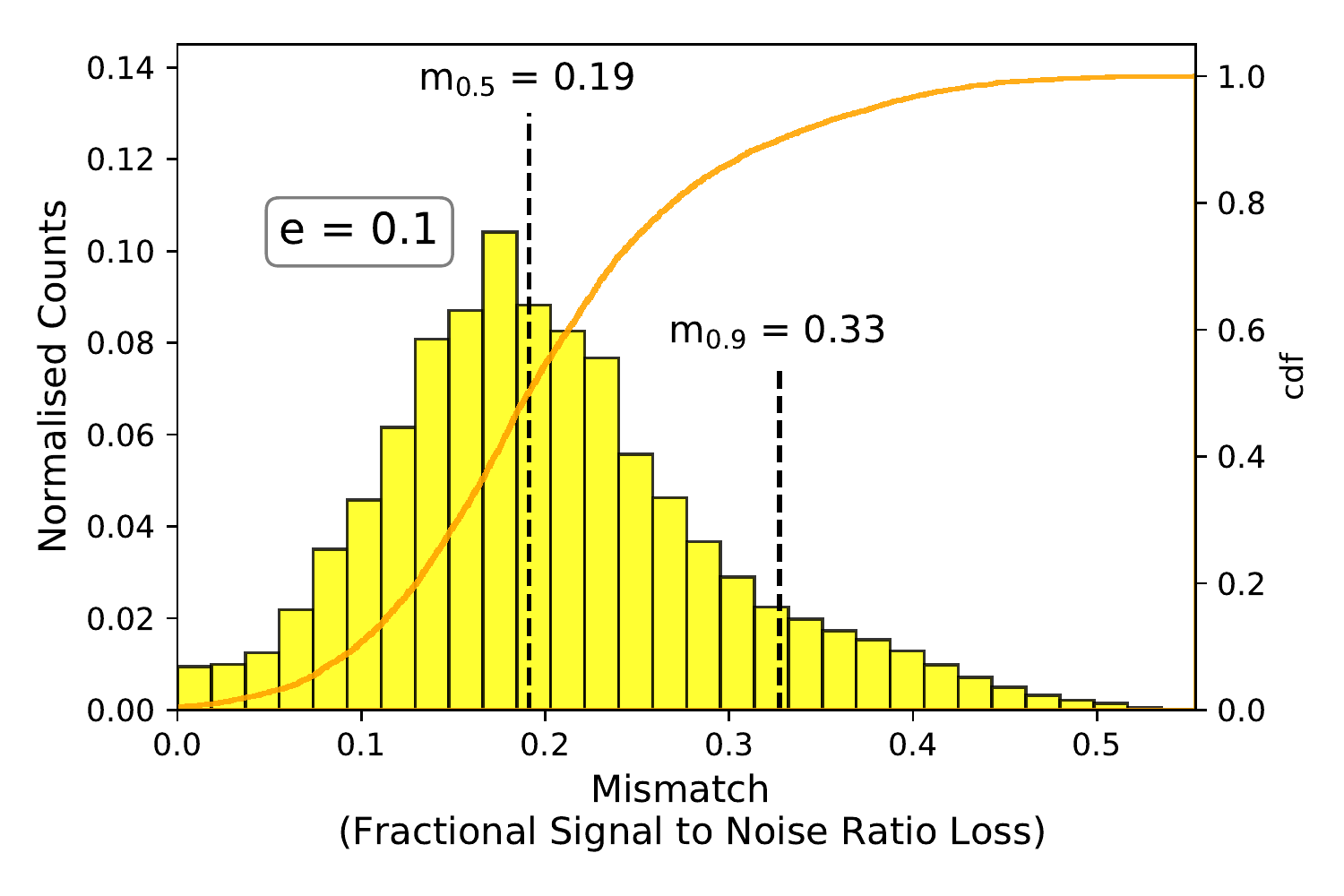}
	\caption{Mismatch distribution of our template-bank for simulated pulsar-black hole binaries. The dotted line indicates the median and 90th percentile mismatch values.}

    \label{fig:verification template bank}
\end{figure}

After constructing the template bank, the next step is to verify that the template bank satisfies the chosen coverage and mismatch criteria. We did this by simulating PSR-SBH binaries in a 72-minute observation with the pulsar rotating at the maximum spin-period chosen for our template bank $P_{\rm spin} = 15$ ms. We simulated 10,000 binaries with a uniform distribution of orbital period between 6-12 hours, a fixed eccentricity of 0.1 and inclination angle varying from 0 to 90 degrees. We let the orbital phase and longitude of periastron vary between a value from 0 to 2$\pi$ radians. All the parameters used in the simulation can be found in table \ref{tab:simulation parameters}. We also simulated an isolated pulsar in order to calculate our best case signal to noise ratio. The mismatch distribution from running our elliptical binary search pipeline is shown in figure \ref{fig:verification template bank}. We created the template bank to have a coverage of $\eta = 90\%$ and a mismatch of 0.2. Since we also search over frequency using an FFT, we need to account for the additional mismatch from this parameter. The response of the FFT in frequency is not uniform. It's maximum for signals falling in the centre of the fourier bin and loses sensitivity as you go towards the edge of a bin. This is called \emph{scalloping} (see for e.g. \cite{1993ispu.conf..372M}). One option to mitigate this is by padding the timeseries with its mean value to minimise the effect of FFT scalloping (See section 3.8.4 of \citet{2011PhDT.......293K} or \citet{2002AJ....124.1788R} for a detailed discussion on this). We avoid this in our analysis in order to minimise computational time. The 90th percentile of our mismatch distribution is 0.33, i.e.\ we cover $90\%$ of our parameter space with a mismatch value $m_{0.9}\leq 0.33$. Therefore taking into account the additional mismatch from the spin-frequency parameter, the requirements of the template-bank have been fulfilled for our chosen search set-up.
\section{Comparing to other pulsar-search pipelines}
\subsection{Tests on Simulated Observations of Mildly Eccentric P-SBH Binaries}
\label{sec:compare}
In this section, we compare the performance of a \textsc{PRESTO} -based pipeline doing an acceleration and jerk search to a template-bank search over three Keplerian parameters assuming a circular orbit and over five Keplerian parameters assuming an elliptical orbit. We restrict our analysis to binaries in the 5-10 $T_{\mathrm{obs}}$ regime. We start by simulating 10,000 PSR-SBH Binaries with an eccentricity of 0.1, companion mass of 8 $M_{\odot}$, pulsar mass of 1.4 $M_{\odot}$ and a spin-period of 15 ms. These values are kept fixed. We simulate orbits with a uniform distribution of orbital period between 6 and 12 hours along with a uniform distribution of inclination angles, orbital phase and longitude of periastron values. We used a modified version of the code \textsc{FAKE} from \textsc{Sigproc}\footnote{\url{https://github.com/SixByNine/sigproc}} for our work. A full list of our input spin and orbital parameters for the simulation can be found in table \ref{tab:simulation parameters}. Additionally, we simulated an isolated pulsar with identical input parameters in order to calculate the best-case detection significance that can be recovered. For each simulated binary pulsar, we run each of our pipelines and calculate the mismatch value using the following expression.
\begin{equation}
m = 1 - \frac{\mathrm{(SNR)_{Recovered,Search Pipeline}}}{\mathrm{(SNR)_{Isolated}}}.
\end{equation}
\textsc{PRESTO}  implements a frequency-domain acceleration/jerk search (FDAS/FDJS). FDAS assumes that the pulsar's acceleration $A$ is constant throughout the observation. Under this assumption, each harmonic of the pulsar signal will experience an acceleration $A$ given as:
\begin{equation}
\label{eq: accel value presto}
    A = \frac{\dot f c}{h f} = \frac{z c}{h f t\mathrm{_{obs}}^2},
\end{equation}
where $z$ is the frequency
derivative in units of fourier bins (i.e. how many bins the signal drifts
during the observation), $h$ is the harmonic number of the signal, $f$ is the spin-frequency and $\dot f$ is the spin-frequency derivative. Similarly, FDJS assumes a constant jerk, $J$, throughout the observation and the previous equation can be expanded to
\begin{equation}
\label{eq: jerk value presto}
    \dot A = J = \frac{\ddot f c}{h f} = \frac{w c}{h f t\mathrm{_{obs}}^3},
\end{equation}
where $w$ corresponds to the number of fourier frequency derivative bins that the signal is smeared onto during the course of an observation. The user can input the maximum number of fourier frequency bins ($z\mathrm{_{max}}$) and/or the maximum number of fourier frequency derivative bins ($w\mathrm{_{max}}$) to search over in an observation.

For FDAS, we used the maximum value allowed by the software $z\mathrm{_{max}}$ = 1200 which for a 15-ms pulsar in a 72-minute observation corresponds to an acceleration of 289.35 ms$^{-2}$ for the fundamental ($h$ = 1). For FDJS, we used a $z\mathrm{_{max}}$ = 600 and a $w\mathrm{_{max}}$ = 1800 which corresponds to an acceleration of 144.68 ms$^{-2}$ and a jerk of 0.1 ms$^{-3}$. The maximum values used for acceleration and jerk in our simulations were limited by software and available RAM. In principle, these simulations could be repeated with a broader range of acceleration and jerk, however we do not expect a significant improvement in performance as the assumption of constant acceleration or jerk throughout the observation breaks down for extreme values.

\begin{figure*}
        \centering
        \begin{subfigure}[b]{0.48\textwidth}
            \centering
            \includegraphics[width=1.1\textwidth]{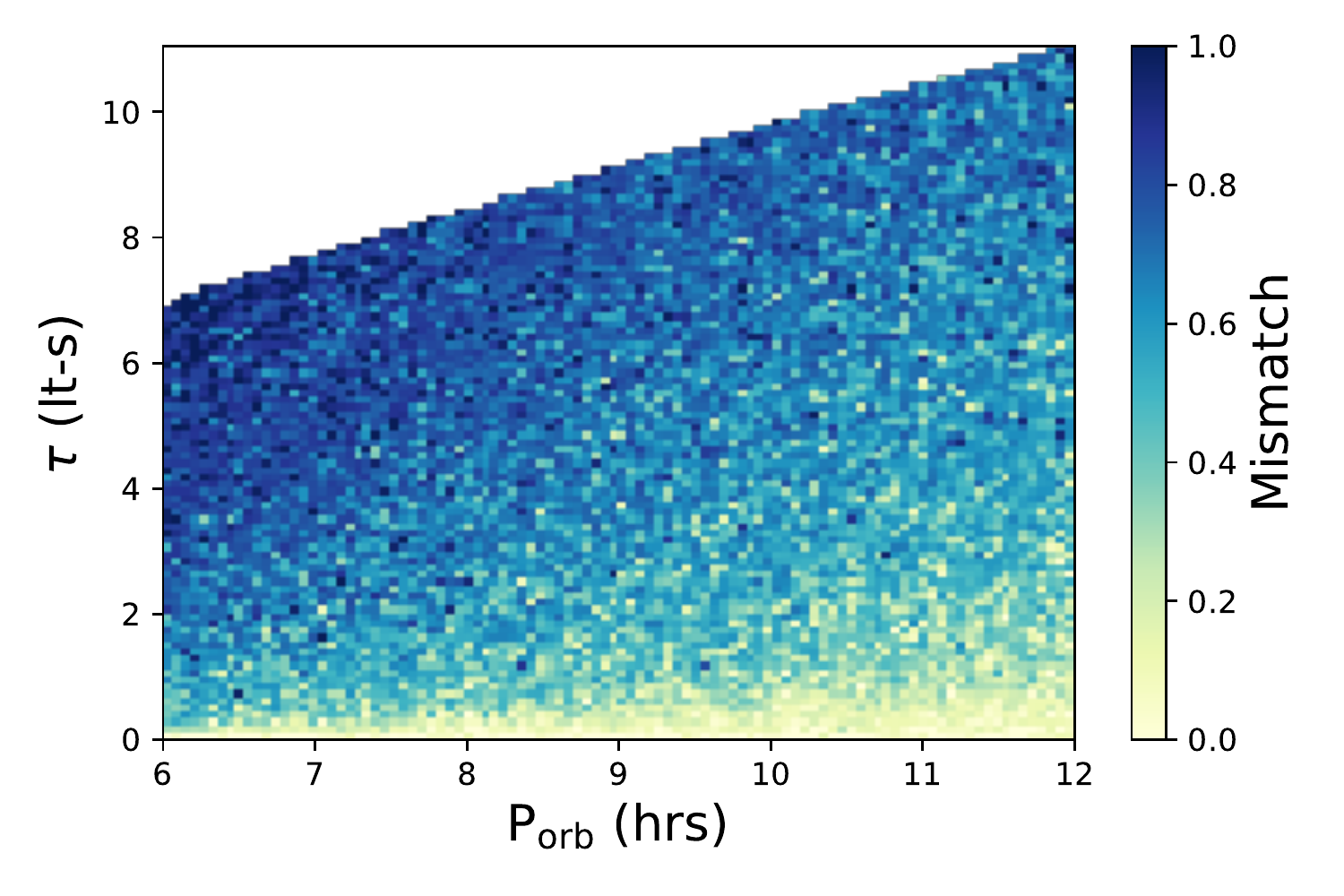}
            \caption{Acceleration Search over f and a (z$_{max}$ = 1200)}
            \label{fig:mean and std of net14}
        \end{subfigure}
        \hfill
        \begin{subfigure}[b]{0.48\textwidth}  
            \centering 
            \includegraphics[width=1.1\textwidth]{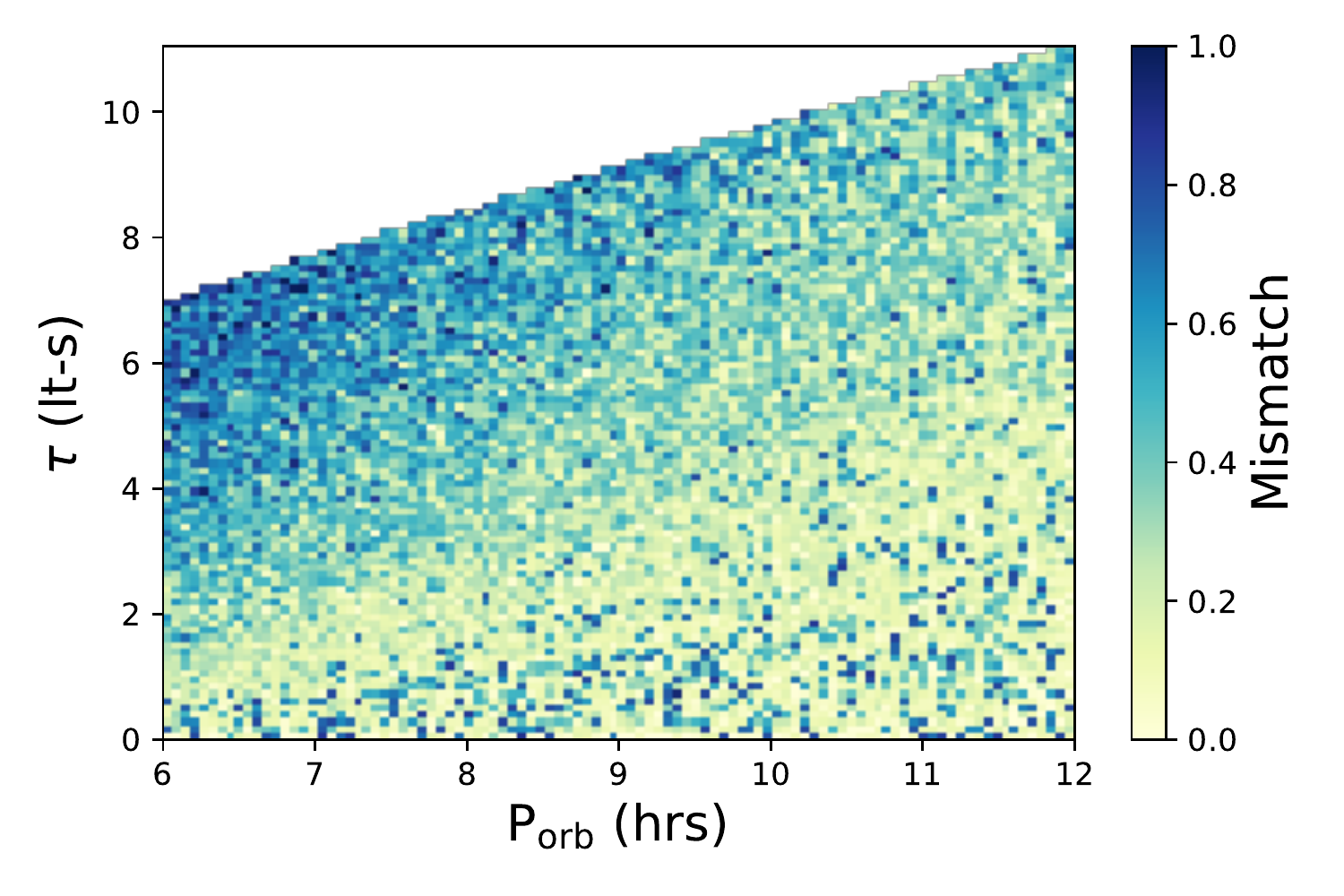}
           \caption{Jerk Search over f, a and j (z$_{max}$ = 600, w$_{max}$ = 1800)}   
            \label{fig:mean and std of net24}
        \end{subfigure}
        \vskip\baselineskip
        \begin{subfigure}[b]{0.48\textwidth}   
            \centering 
            \includegraphics[width=1.1\textwidth]{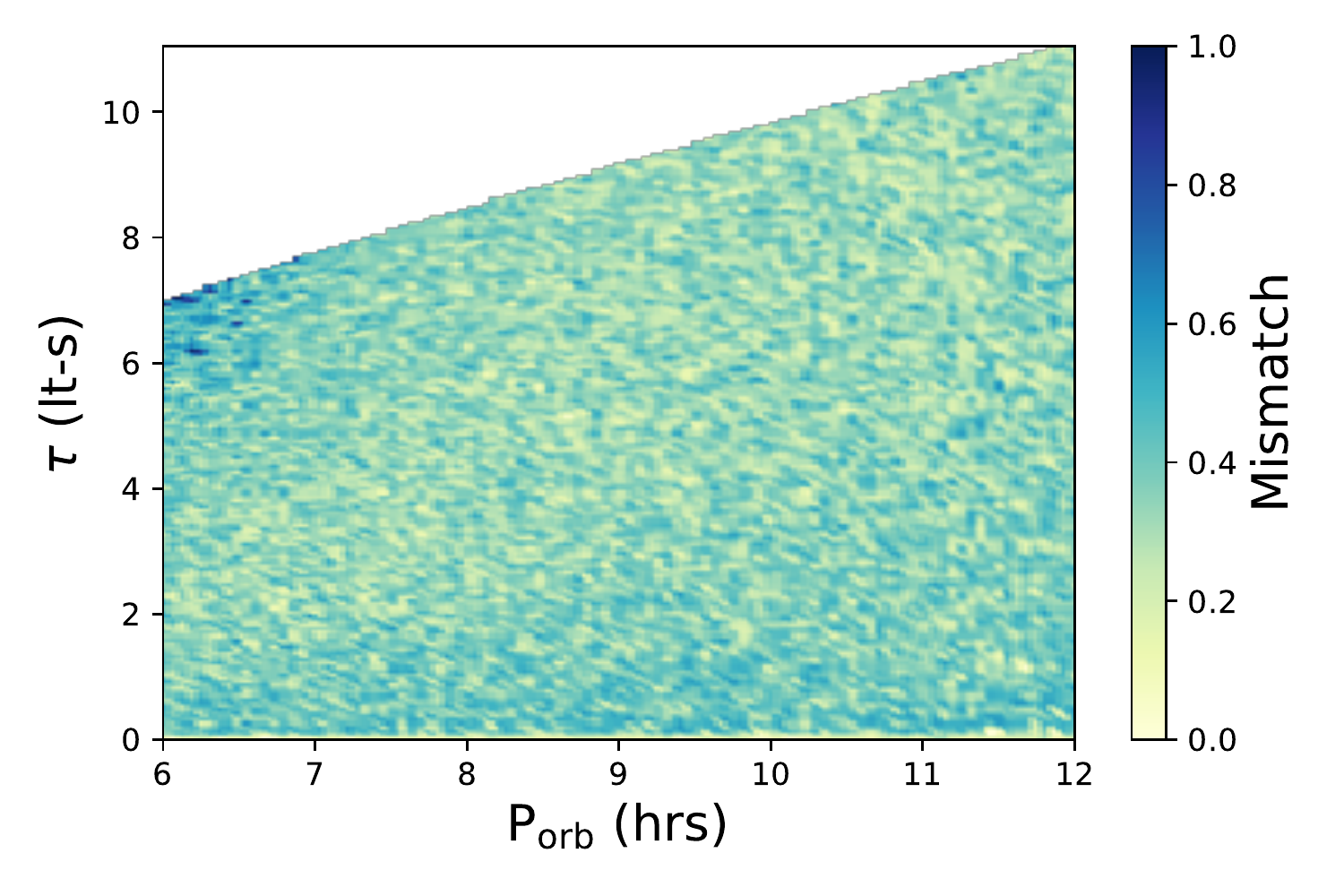}
            \caption{Template-Bank Search over f and Three Circular Orbit Keplerian Parameters.}  
            \label{fig:mean and std of net34}
        \end{subfigure}
        \hfill
        \begin{subfigure}[b]{0.48\textwidth}   
            \centering 
            \includegraphics[width=1.1\textwidth]{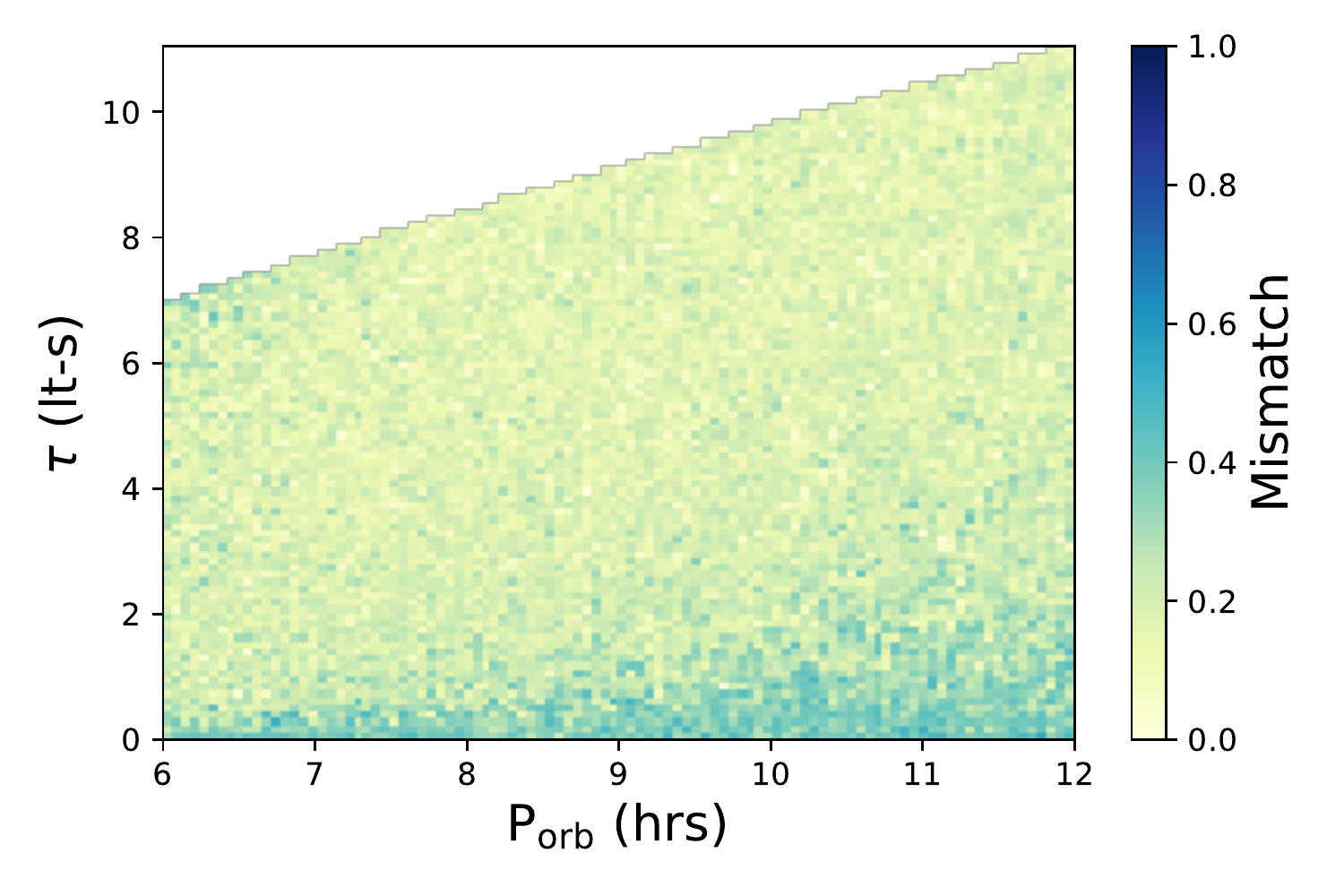}
           \caption{Template-Bank Search over f and Five Elliptical Orbit Keplerian Parameters.}  
               
            \label{fig:mean and std of net44}
        \end{subfigure}
        
        \caption{Recovered signal-to-noise ratio expressed in terms of mismatch using different pulsar-search pipelines for simulated PSR-SBH binaries with eccentricity 0.1. Mismatch values shown here are averaged across different values of initial orbital phase and longitude of periastron. These plots contain a grid with 10,000 points. In each point, we simulate a binary, and report the recovered mismatch. As expected, we notice a high loss in signal to noise ratio for short orbital periods for an acceleration search. We notice a significant improvement with a jerk search, however it still fails to recover signals for short orbits and high a$\sin \epsilon$ values. We notice a further improvement by searching over three Keplerian parameters with an average mismatch value of 0.35 throughout the entire parameter space. The average mismatch value by searching over all five Keplerian parameters is 0.18. The cloudy structure in the template-bank search plots come from the fact that we have a probabilistic coverage of the parameter space $\eta$ = 0.9, therefore for some regions we have a higher mismatch than the nominal value.} 
        \label{fig:compare all algos}
    \end{figure*}

\begin{figure*}
	\includegraphics[width=2.3\columnwidth]{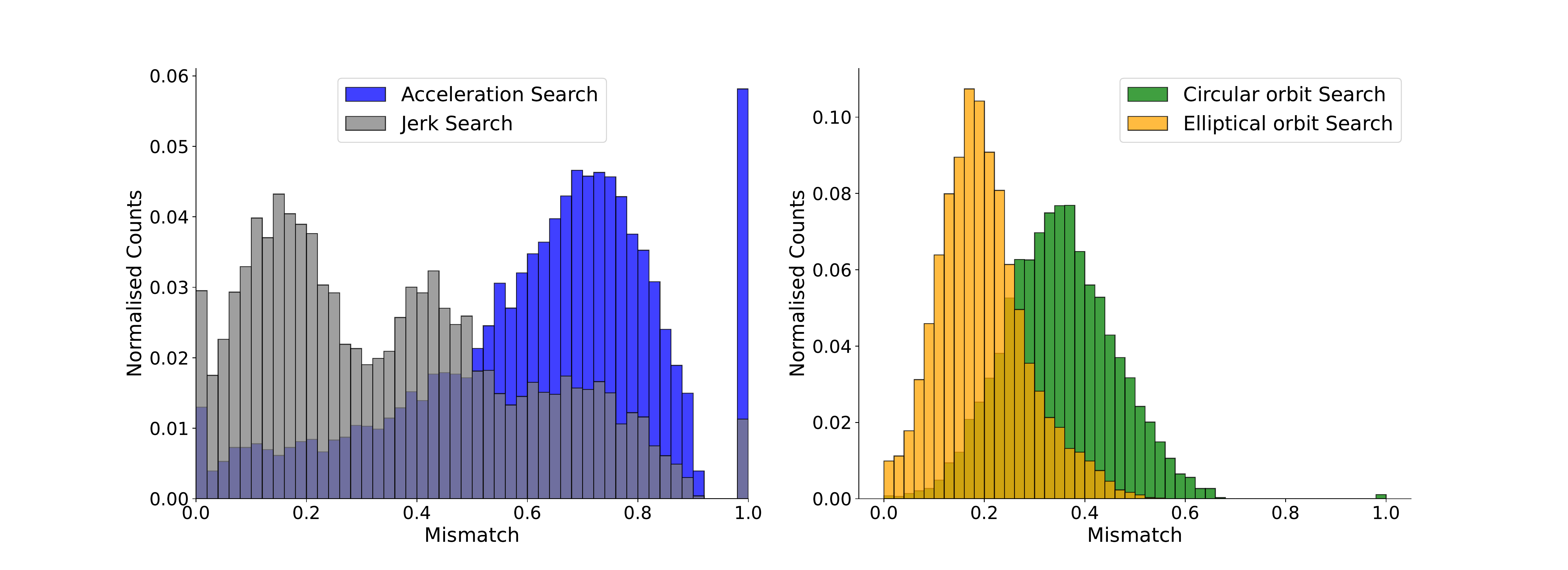}
    \caption{1-D histogram of fractional signal to noise recovered in terms of mismatch values from a polynomial based acceleration and jerk search (left) compared to Keplerian searches (right) assuming a circular orbit binary and an elliptical orbit binary. These simulations were carried out on 72-minute observations of mildly recycled (15 ms) PSR-SBH binaries ($m\mathrm{_{companion}}$ = 8 $\mathrm{M_{\odot}}$) with an orbital period range of 6-12 hours in eccentric orbits (e=0.1).}
    \label{fig:histogram compare}
\end{figure*}

For the template-bank search, we implemented two different searches. One assumes a circular orbit binary and searches over the three Keplerian parameters ($P\mathrm{_{orb}}$, $\tau$, $\alpha$) and our second search expands the parameter space to include eccentricity $e$ and longitude of periastron ($\psi$). For the circular search, we generated 10,128 orbital templates using the random template-bank algorithm with the same binary parameter range as our simulations but with $e = 0$. For the elliptical search, we generated 2,04,7716 orbital templates. Both these values were computed from equation \ref{eq:total templates} assuming a coverage $\eta = 0.9$ and mismatch $m = 0.2$. Our results are shown in figure \ref{fig:compare all algos}. These plots were made on a $100 \times 100$ grid. At each point, we simulated a binary pulsar and report the recovered mismatch value from each pipeline. The plots shows the mismatch value averaged across initial orbital phase and longitude of periastron. We notice a general trend of loss of signal-to-noise ratio (high mismatch values) using polynomial based searches for binaries with short orbital periods. Additionally, a high $a\sin{\epsilon}$ value resulting from a steeper inclination angle also makes them harder to detect. For the special case of $a\sin{\epsilon}$ = 0 (face-on) orbit, the Doppler modulation is negligible, hence we notice a low mismatch value from all pipelines. Jerk search provides a significant improvement over acceleration search. However, this approach also loses sensitivity for binary pulsars with short orbital period binaries and high $a\sin{\epsilon}$ values. The coherent search over three Keplerian parameters provides a uniform sensitivity throughout our parameter space with an average mismatch value of 0.35. Depending on the flux density of the pulsar, this value may or may not be acceptable. As expected, we notice the most improvement by searching over all five Keplerian parameters. We notice an average mismatch value of 0.19 for pulsars in our parameter space. The 1-D histogram for the mismatch values for all the pipelines can be seen in figure \ref{fig:histogram compare}. Additionally, we also report the median and 90th percentile of our mismatch distribution along with the GPU computation time it takes to search one simulated binary in table \ref{tab:median}. For computational benchmarks, we used the GPU accelerated version of \textsc{presto}\footnote{\url{https://github.com/jintaoluo/presto_on_gpu}}.

\begin{table*}
\begin{center}
\caption{Median and 90th percentile of mismatch values from our simulations for different pulsar search pipelines. We also report here the associated search parameters for each pipeline and the compute time per simulation. These numbers were benchmarked on a Nvidia Quadro RTX 6000 GPU and a Intel(R) Xeon(R) Platinum 8268 CPU. We used the GPU accelerated version of \textsc{presto} for the acceleration search. }
\label{tab:median}

\begin{threeparttable}
\begin{tabular}{cccc}
\toprule
Pipeline & m$_{50}$ & m$_{90}$ & GPU Compute time to search one simulated binary (mins) \\

\midrule
Acceleration Search (z$_{max}$ = 1200) & 0.66 & 0.86 & 0.4 \\
Jerk Search (z$_{max}$ = 600, w$_{max}$ = 1800) & 0.35 & 0.73 & 840.1 (32.9)\tnote{a} \\
Circular Orbit Search ($\eta$ = 0.9, m = 0.2) & 0.35 & 0.49 &  1.2  \\
Elliptical Orbit Search ($\eta$ = 0.9, m = 0.2) & 0.19 & 0.33 &  990.0 \\
 
\hline

\end{tabular}
\begin{tablenotes}
  \item[a] The value outside the parenthesis is the time taken for running jerk search on a CPU. An open-source GPU accelerated Fourier domain search jerk search pipeline currently does not exist. The value inside the parenthesis is the approximate time a GPU version of the pipeline would take assuming same speed factor of 25.5 which was observed between the CPU and GPU version of acceleration search. 
  \end{tablenotes}
\end{threeparttable}
\end{center}
\end{table*}

\subsection{Tests on the Double PSR~J0737-3039}
\subsubsection{Comparison to an Acceleration and Jerk-Search \textsc{presto}-based pipeline}
In this section, we do a similar comparison of the performance of different pulsar search pipelines on an archival Parkes observation of the double pulsar PSR~J0737-3039 that covers one full orbit of the system. The original observation had an integration time of 2.6 hours with a time sampling interval of \SI{80}{\micro \second}. The data was dedispersed at DM~=~\SI{48.92} {\parsec \per \centi \meter \cubed} to form a timeseries. We then downsampled the data by a factor of 16 to \SI{1280}{\micro \second}. We compared all the pipelines on two different integration lengths of 90-minute and 45-minute duration. Additionally, we also searched the data by shifting the starting point of the observation in steps of 100s. This effectively helped us sample different orbital phases of PSR~J0737-3039 and test its detectability using different pipelines. 

\begin{figure*}
	\includegraphics[width=2.2\columnwidth]{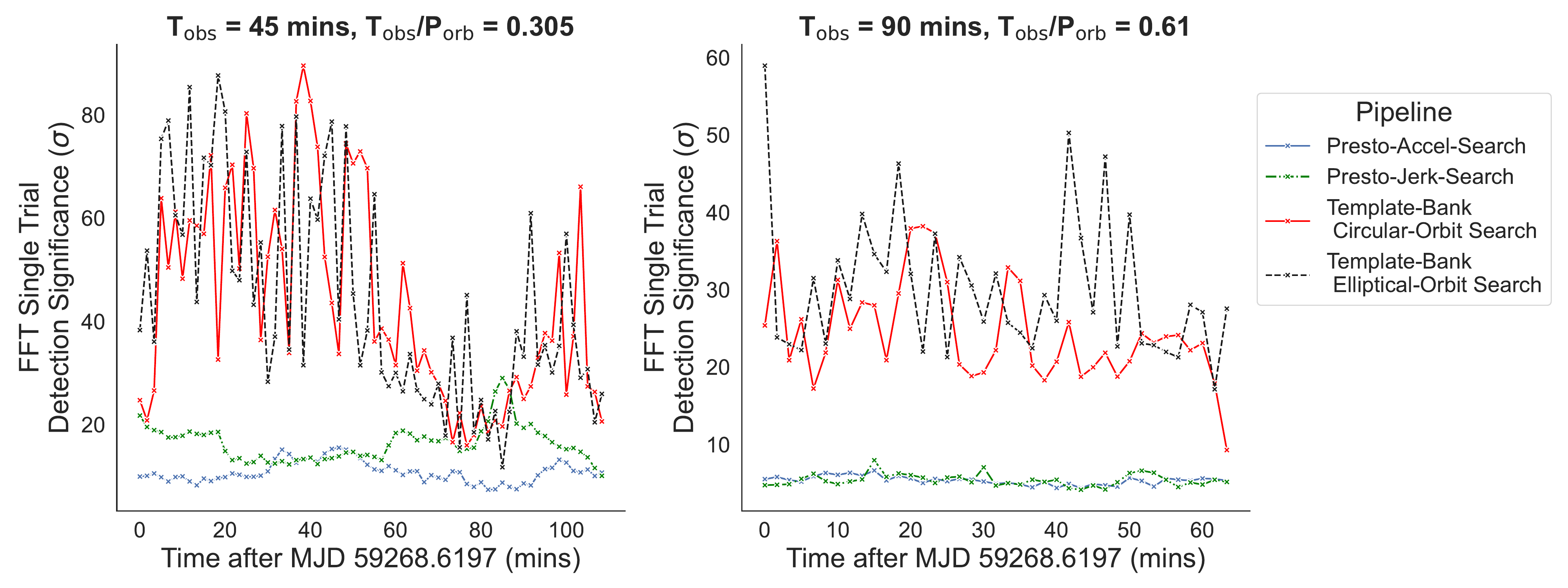}
	\includegraphics[width=2.2\columnwidth]{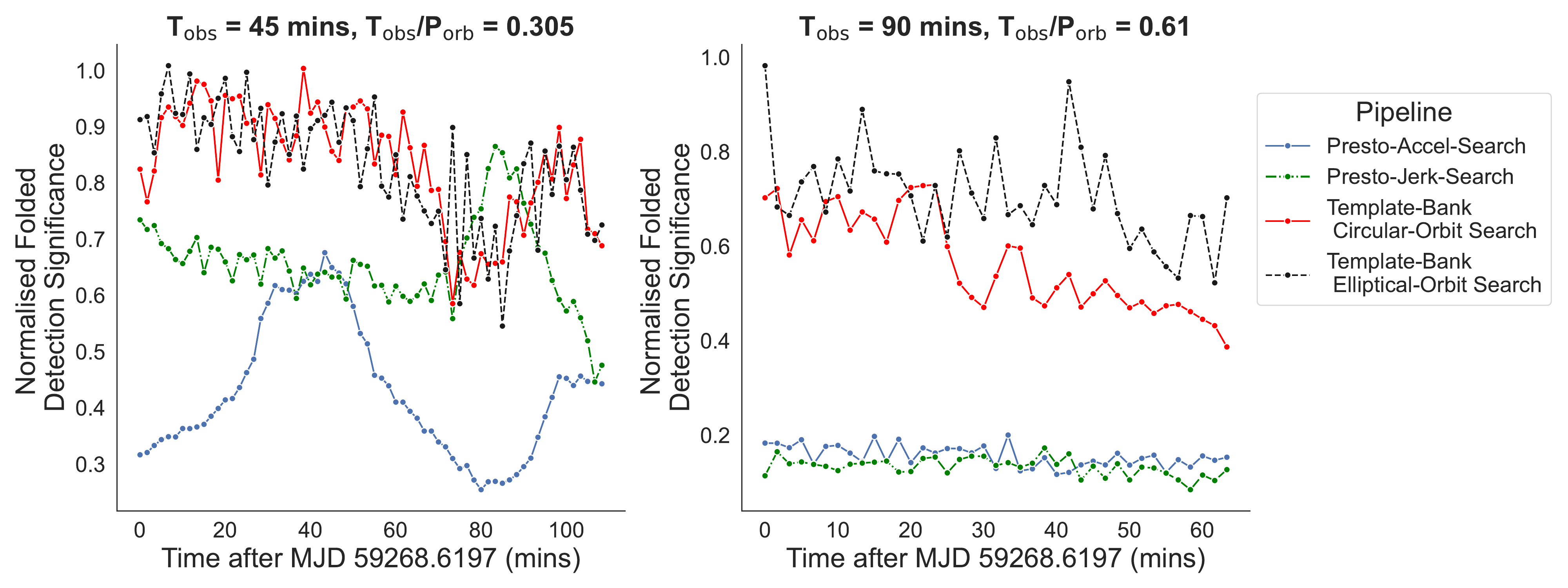}
    \caption{Results from running a frequency domain \textsc{presto}-based acceleration, jerk search pipeline and a template bank pipeline assuming a circular and elliptical orbit on different integration lengths across a 2.6 hour Parkes observation of PSR~J0737-3039A. Left panel shows results from searching over 45-minute observations and right panel shows results from searching over 90-minute observations. The top row shows the FFT single-trial detection significance calculated from the sifting routine as a function of starting point of the observation incremented by 100 seconds. The bottom row shows the  folded detection significance normalised by the detection significance obtained from folding the ephemeris of the pulsar. These values were obtained from \textsc{PREPFOLD}. Both these routines are available in the \textsc{presto} software package.}
    \label{fig:compare J0737}
\end{figure*}

For our searches using the \textsc{PRESTO} pipeline we used $z_{max} = 1200$ for acceleration search which using equation \ref{eq: accel value presto} corresponds to a maximum acceleration value of \SI{1120.95} {\meter \per \square \second} for the fundamental in the 45-minute observation  and \SI{280.24} {\meter \per \square \second} in the 90-minute observation. Similarly, for jerk search we used a $z_{max} = 600$ and $w_{max} = 1800$ which using equation \ref{eq: jerk value presto} corresponds to maximum acceleration of \SI{560.47} {\meter \per \square \second} and jerk \SI{0.62} {\meter \per \cubic \second} in the 45-minute observation. For the 90-minute observation, this corresponds to a maximum acceleration of \SI{140.2} {\meter \per \square \second} and a maximum jerk of \SI{0.078} {\meter \per \cubic \second}.  As mentioned in the previous section, our jerk search range was limited by the amount of available RAM in the system. In order to get around this, in the next section we also compare our results to a \textsc{sigproc}-based time-domain acceleration and jerk search pipeline where we search over a much wider acceleration and jerk search range to recover the best-case sensitivity from a first and second order polynomial based search pipeline. For the elliptical search template-bank, we set the maximum eccentricity to be equal to 0.08 and longitude of periastron was set to vary from 0-2$\pi$. For all the template-banks, the minimum pulsar mass and maximum companion mass was kept fixed at $1.4 M_{\odot}$ and 1.6 $M_{\odot}$ respectively. We also searched over all possible values of initial orbital phase and inclination angles. We tweaked the coverage, mismatch, spin-period and orbital period values for different template banks in order to do a feasible search. These values are shown in table \ref{tab:template bank j0737} and our results are shown in figure \ref{fig:compare J0737}. 

In order to have a fair comparison between different pipelines, we present the single-trial detection significance of each pipeline. In the top panel of figure \ref{fig:compare J0737}, we present the single-trial FFT detection significance in the y-axis which is the sigma value reported from the sifting routine\footnote{\url{https://github.com/scottransom/presto/blob/master/examplescripts/ACCEL_sift.py}} in \textsc{PRESTO}. The sigma reported is the probability that a given signal might be due to noise, but expressed in terms of equivalent Gaussian sigma. For the detections from the template-bank pipeline, we applied a time-domain resampling scheme (explained in section \ref{sec:time domain resampling}) on the dedispersed timeseries using the orbital template that gave us highest significance detection in the FFT. We then ran the code \textsc{accelsearch} from \textsc{PRESTO} with a $z_{max} = 4$\footnote{We also tested the detection significance with a $z_{max} = 2$ and got identical results.} and compared the sigma value reported to those obtained from the standard acceleration and jerk search. 

The x-axis shows the starting point of our observation. Our results show that for 45-minute observations, acceleration search can partially recover the signal for some orbital phases but the detection is relatively weak for most of the orbital phase. As expected, jerk search improves the detection significance substantially. However, its sensitivity is still worse compared to full orbital searches. We don't notice a significant difference between the detection significance from elliptical and circular orbit searches. Using the Kolmogorov-Smirnov (KS) test, we obtain a p value of 0.99 and therefore the null hypothesis (the distribution of detection significance from circular orbit searches is identical to that of the distribution of detection significance from elliptical orbit searches) in this case cannot be ruled out. This is likely due to the fact that for detecting Double-Neutron Star systems for short observations in the $T\rm _{obs}\rm/P\rm _{orb}$ = 30\% range, a five-parameter search is not needed.

For the 90-minute observations, we observe a significant difference in the performance of each pipeline. We notice the limitation posed by the assumption of constant acceleration and/or jerk in a long observation which hinders detection. Here, we also see the improved sensitivity provided by the elliptical search pipeline compared to a circular search. Here using the same KS test as above, we obtained a p-value of 0.0125 between the results from the circular and elliptical orbit searches which makes the improvement statistically significant and therefore in this case the null hypothesis can be ruled out. We would like to stress a limitation of this comparison. Our range for acceleration and jerk search values used in the 90-minute observation was limited by the restrictions imposed by the hardware. We expect an improvement in performance with a wider Acceleration and Jerk search range, however we still expect it to perform significantly worse than full orbital searches due to the varying jerk during an observation. We expand more on this in the next section.

In the bottom panel of figure \ref{fig:compare J0737}, we report the normalised folded significance calculated from the code \textsc{PREPFOLD} from \textsc{PRESTO}  as a function of the starting point of the observation. For comparison, we normalised the significance value from each pipeline by dividing it by the significance obtained from folding the ephemeris of the pulsar. The significance reported here for all search pipelines also assumes a single-trial search and therefore can be directly compared. As is this case for FFT detection significance, the  performance of the folded detection significance between circular and elliptical orbit searches is nearly identical for 45 minute observations, but for 90 minute observations, we see a clear statistically significant improvement. Using a KS test, we obtained a p value of 0.99 and 4.6 $\times$ 10$^{-6}$. between the folded detection significance distributions of circular and elliptical search pipeline for 45-minute and 90-minute observations respectively. In a practical search, the value of the folded significance is significant as candidates are thresholded for inspection based on these values. The main advantage of the Keplerian search pipelines compared to polynomial-based search pipelines is its higher sensitivity in longer observations and additionally we need not rely on favorable orbital phases for detection. The non-uniform sensitivity of the template-bank for different orbital phases is due to our probabilistic coverage ($\eta$) of the parameter space described earlier.

\begin{table*}
\caption{Orbital Parameters used to generate the template bank for tests on the observation of PSR~J0737-3039A.}
\label{tab:template bank j0737}
\begin{center}
\begin{threeparttable}
\begin{tabular}{lrrrr}
\toprule
Input Parameter & \multicolumn{2}{c}{$T_\text{obs = 45min}$} & \multicolumn{2}{c}{$T_\text{obs = 90min }$}  \\
& Circular Search & Elliptical Search & Circular Search & Elliptical Search  \\
\midrule
Coverage ($\eta$) & 0.9 & 0.9 & 0.9 & 0.6 \\
Mismatch (m) & 0.05 & 0.05 & 0.2 & 0.6 \\
Max.Spin Period (P$\mathrm{_{spin}}, \rm ms$) & 20 & 22 & 22 & 22\\
Orbital Period range ($\mathrm{P_{orb}, \rm hrs}$) & (2, 7.5) & (2.450-2.457) & (2.33-15) & (2.450-2.456)\\
No. of trials & 65,489 & 52,981 & 615,800 & 16,92, 0617\\
\bottomrule
\end{tabular}

  \end{threeparttable}
\end{center}
\end{table*}

\subsubsection{Comparison to an Acceleration and Jerk-Search \textsc{sigproc}-based pipeline}
As mentioned in the previous section, our jerk search range in the \textsc{presto}-based pipeline was limited by the available RAM in our system. In order to get around this, here we compare our results with a \textsc{sigproc}-based time-domain acceleration and jerk-search pipeline. For both the 45 and 90-minute observations, we searched over an acceleration range of (-500, +500) \SI{}{\meter \per \square \second} and jerk range of (-20,~+20)~\SI{}{\meter \per \second \cubed}. Our step sizes were chosen based on the recommendations given in \citet{2013MNRAS.431..292E}. We generated a total of 295 acceleration trials and 27 jerk search trials in the 45 minute observation case and 1177 acceleration and 211 jerk search trials to search over the 90 minute observations.

\begin{figure*}
	\includegraphics[width=2.2\columnwidth]{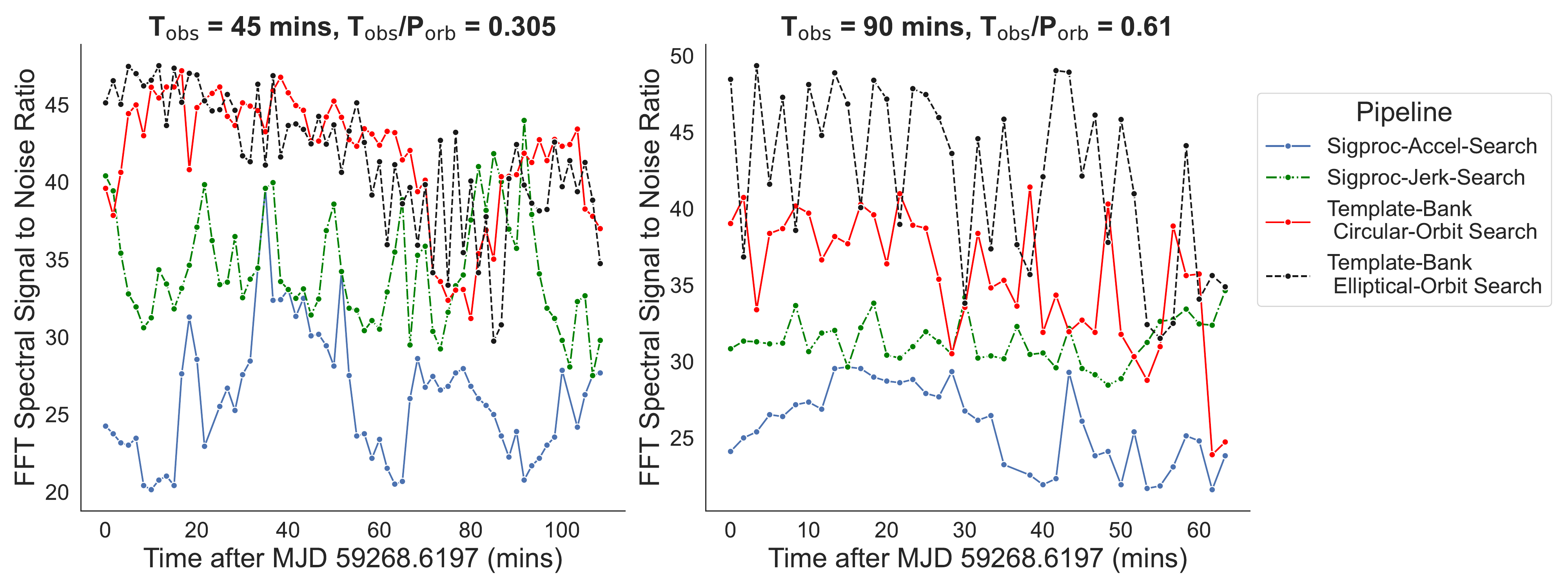}
    \caption{Results from running a time-domain \textsc{sigproc-based} acceleration, jerk search pipeline along with a template bank pipeline assuming a circular and elliptical orbit binary on different integration lengths across a 2.6 hour Parkes observation of PSR~J0737-3039A. Left panel shows results from searching over 45-minute observations and right panel shows results from searching over 90-minute observations. Here we report the FFT spectral signal to noise ratio as a function of the starting point of the observation. The improvement in the sensitivity of the acceleration and jerk-search pipeline compared to the results from the \textsc{presto} pipeline is due to our wider acceleration and jerk search range in the 90-minute observations.}
    \label{fig:compare J0737 sigproc}
\end{figure*}

We used the program \textsc{seek} in \textsc{sigproc} to do the resampling and the FFT which then returns a list of the top candidates. For the results from the template-bank pipeline, we apply our own time-domain resampling code and then pass this resampled timeseries to \textsc{seek}. All the top candidates from each trial are then accumulated and sifted through the software \textsc{PulsarHunter best} (ph-best)\footnote{\url{https://github.com/SixByNine/pulsarhunter}} which removes duplicate detections keeping only the candidate with the highest FFT spectral signal to noise ratio (SNR). In figure \ref{fig:compare J0737 sigproc}, we report this recovered SNR as a function of the starting point of the observation for each pipeline. For the 45 minute observations, we observe consistent results with jerk search offering an improvement over an acceleration search and the results between the circular and elliptical orbit pipelines being nearly identical. For the 90 minute observations, we get improved results from both the acceleration and jerk search pipeline compared to the \textsc{presto}-based pipeline due to our wider acceleration and jerk search range. As expected, we noticed the most improvement by using the template-bank pipeline assuming an elliptical orbit binary followed by the circular orbit search. We would like to stress here that the aim of this exercise was not to do a comparative analysis between a frequency-domain and time domain search pipeline but to compare the best-case results from an acceleration and jerk search pipeline to a full orbital search. 

\section{Scaling of Required Orbital Templates as a function of each Keplerian Parameter}
\label{sec:scaling relation keplerian}
In this section, we describe the scaling relation of the total required random templates for a search when we change one of the Keplerian parameters while leaving the rest unchanged. From equation \ref{eq:total templates}, we know that for a given coverage, mismatch and dimension of a search (n = 3 for circular orbit searches, and n = 5 for elliptical orbit searches), the total number of required random templates is directly proportional to the volume of the parameter space. Therefore, our scaling relations are calculated by tuning one Keplerian parameter at a time and comparing it to the changes observed in the proper volume of the parameter space (equation \ref{eq:volume}). 

To compute the volume, we need to specify the maxima and minima for each parameter in our search range. For all the results, described here, we assumed a $P_{\rm orb, max} = 10 T_{\rm obs}$ and initial orbital phase ($\alpha$) between 0 and 2 $\pi$. Additionally, for the elliptical orbit searches, we also integrate across the longitude of periastron parameter ($\psi$) between 0 and 2 $\pi$. We present here rough scaling relations of the total required templates versus observation time (T$_{\rm obs}$), minimum orbital period ($P_{\rm orb, min}$), maximum companion mass (M$_{\rm c, max}$), projected radius ($a\sin{\epsilon}$) and additionally maximum eccentricity $e_{\rm max}$ for the elliptical orbit searches.

The templates scaling relation depends on the total number of orbits visible in an observation. In most radio pulsar searches that use an acceleration or a jerk search, we are interested in the regime where only a fraction of the orbit is visible. An alternative technique is sideband searches \citep{2003ApJ...589..911R} which is used to cover the parameter space when the observation time is longer than the orbital period. Here, we present scaling relations that cover the range 0.1-5 i.e from when 10 \% of the orbit is visible in an observation to up-to 5 full orbits. Here we define number of orbits with respect to the minimum orbital period covered in the search. To calculate the volume integral (equation \ref{eq:volume}), we used a monte-carlo integration technique  which gives us the proper volume and the associated volume error estimate. Since the templates scaling relation generally follow a power law, we transform them into log units and perform a weighted linear regression\footnote{Each point was weighted by the reciprocal of the variance of the proper volume estimate in log units.} in order to calculate the scaling relation. 
\subsection{Scaling for Circular Orbit Searches}
\begin{figure*}
	\includegraphics[width=2.0\columnwidth]{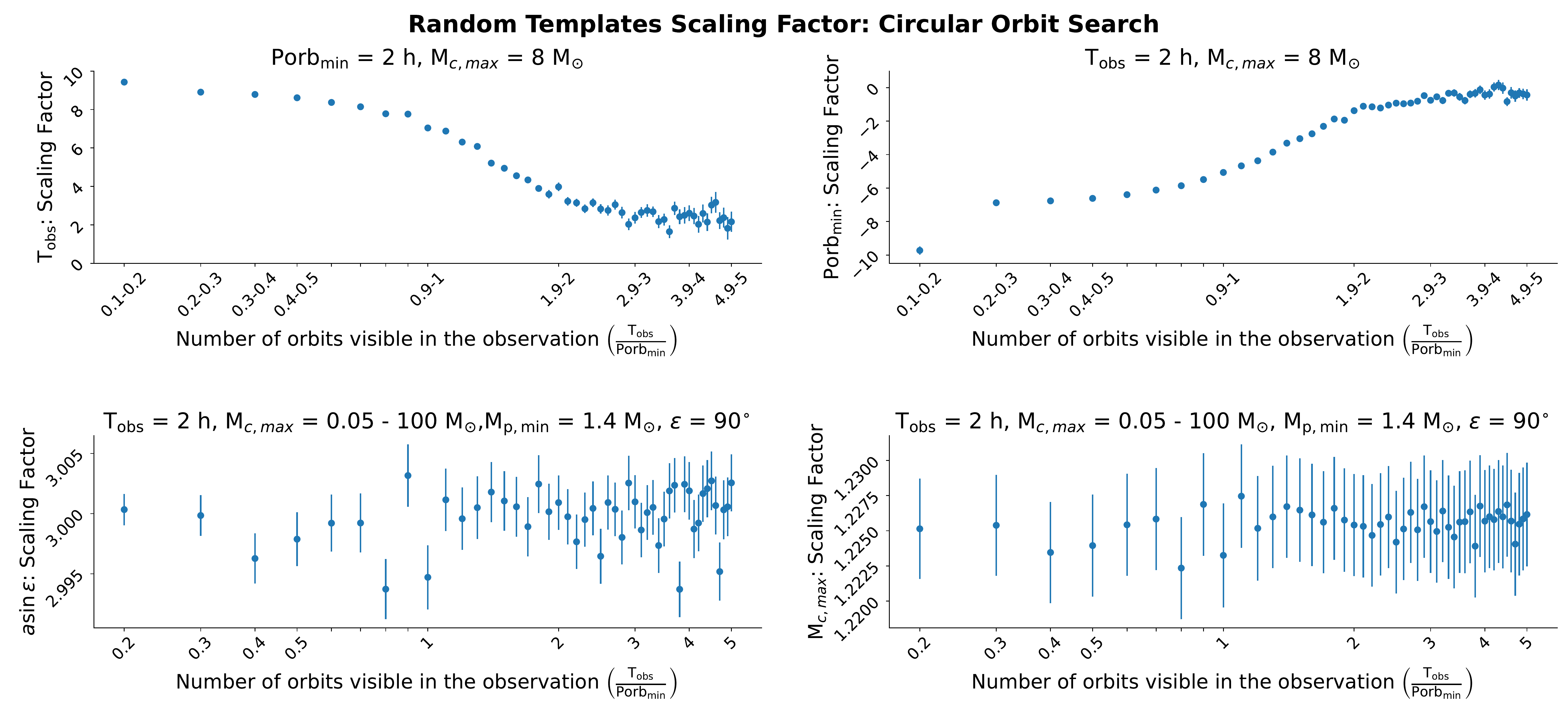}
    \caption{Scaling relation for the total number of required random templates for a circular orbit search calculated by varying one search parameter and keeping the rest fixed. In the top left we show the scaling relation of observation time (T$_{\rm obs}$) versus the number of orbits visible in the observation. Top right is the scaling relation for minimum orbital period (Porb$_{\rm min}$). Bottom row shows the scaling relation for projected radius ($a\sin{\epsilon}$) and maximum companion mass (M$_{\rm c, max}$) respectively. These were computed by estimating the proper volume of the parameter space. See main text for further discussion.}
    \label{fig:Scaling relation circular}
\end{figure*}

Our scaling relations for circular orbit searches are presented in figure \ref{fig:Scaling relation circular}. In the top left, we present the scaling relation for observation time versus the number of orbits visible in an observation.\footnote{defined as the ratio between observation time and minimum orbital period} Here, we have kept the minimum orbital period fixed at $P_{\rm orb, min}={2 \rm h}$ and maximum companion mass M$_{\rm comp, max} =  8 M_{\odot}$. We then tune the observation time to sample points in the parameter space where 0.1-5 orbits are visible. In the top right, we present the scaling relation for minimum orbital period where we do the converse by tuning the orbital period, keeping the observation time fixed at $T_{\rm obs}={2 \rm h}$ and same companion mass. The steepest dependence of templates versus T$_{\rm obs}$ and $P_{\rm orb, min}$ occurs when only a fraction of the orbit is visible. Integrating from 10~\% of the orbit to 20 \% of the orbit, we find that the templates scale to approximately the ninth power for both T$_{\rm obs}^{9.42(1)}$ and $P_{\rm orb, min}^{-9.7(2)}$ where the values in the parenthesis represent our uncertainty in the final digit. When the observation samples larger fractions of the orbit, this dependence flattens out. When two or more orbits are visible in an observation the total templates scale as $\sim$ T$_{\rm obs}^{2}$ and approximately linearly for minimum orbital period $P_{\rm orb, min}^{-1.1(1) - 0.1(3)}$. In the bottom part of figure \ref{fig:Scaling relation circular}, we show the scaling relation for projected radius (a$\sin{\epsilon}$) and maximum companion mass (M$\rm _c, max$). The templates scale as $\sim$ (a$\sin{\epsilon})^3$ and $\rm M_{c, max}^{1.2}$ and does not depend on the number of orbits visible in an observation. For these results, we kept the minimum pulsar mass fixed at M$_{\rm p, min}$~=~1.4 M$_{\odot}$ and varied the maximum companion mass between M$_{\rm c, max}$~=~0.05 - 100 M$_{\odot}$. We assumed an inclination angle $\epsilon$ = 90$^{\circ}$ for our conversion between M$_{\rm c, max}$ and a$\sin{\epsilon}$.

\subsection{Scaling for Elliptical Orbit Searches}
\begin{figure*}
	\includegraphics[width=2.0\columnwidth]{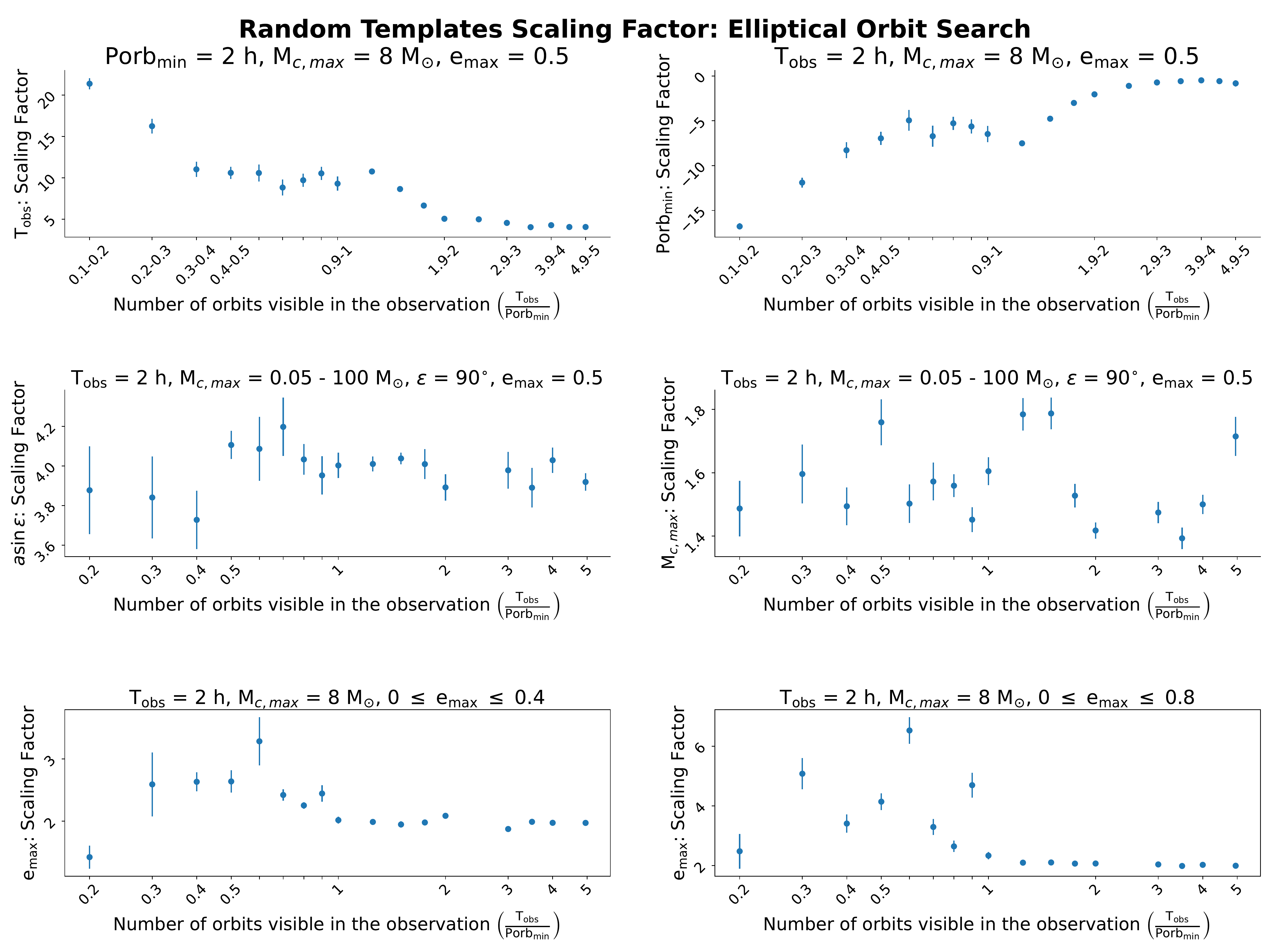}
    \caption{Scaling relation for the total number of required random templates for an elliptical orbit search calculated by varying one search parameter and keeping the rest fixed. In the top left we show the scaling relation of observation time (T$_{\rm obs}$) versus the number of orbits visible in the observation. Here, we have fixed the maximum eccentricity $e_{\rm max} = 0.5$ and maximum companion mass M$_{\rm comp} = 8 M_{\odot}$.  Top right is the scaling relation for minimum orbital period (Porb$_{\rm min}$). Second row shows the scaling relation for projected radius ($a\sin{\epsilon}$) and maximum companion mass (M$_{\rm c, max}$) respectively. Bottom row shows the scaling relation with respect to maximum eccentricity of the search. These were computed by estimating the proper volume of the parameter space. See main text for further discussion.}
    \label{fig:Scaling relation elliptical}
\end{figure*}

We follow the same procedure to calculate the scaling relation for elliptical orbit searches. In addition to the parameters mentioned above, we now have an extra parameter - maximum eccentricity in the search. Our scaling relations are shown in figure \ref{fig:Scaling relation elliptical}. In the top row, we show the results for observation time and minimum orbital period. Here we have the kept the maximum eccentricity fixed at e$_{\rm max}$ = 0.5 and same companion mass of 8 M$_{\odot}$. When only a fraction of the orbit is visible, the dependence of templates on observation time and minimum orbital period is even more steeper for elliptical orbit searches compared to circular orbit searches with the templates scaling as T$_{\rm obs}^{21.3(6)}$ and $P_{\rm orb, min}^{-16.7(3)}$. When one full orbit is visible in an observation the templates scale approximately as $\sim \rm T_{\rm obs}^{10.7(2)}$ and $\sim \rm Porb_{\rm min}^{-6.4(9)}$. When two full orbits are visible, templates scale to the fifth power of observation time $\sim \rm T_{\rm obs}^{5}$ and from 3-5 orbits it scales approximately to the fourth power of the observation time. For the minimum orbital period, we notice that the templates scale approximately linearly when more than two orbits are visible $\sim \rm Porb_{\rm min}^{-1}$. 

In terms of scaling relation for projected radius a$\sin{\epsilon}$ shown in the second row of figure \ref{fig:Scaling relation elliptical}, the templates scale approximately to the fourth power $\sim \rm (a\sin{\epsilon})^4$ and does not significantly change with the number of orbits in the observation. Since the pulsar mass and inclination angle is fixed, the relation for maximum companion masses is similar to a$\sin{\epsilon}$ with the templates scaling approximately to $\sim \rm M_{c, max}^{1.6}$.

Finally, we have the maximum eccentricity parameter for which the results are shown in the bottom row of figure \ref{fig:Scaling relation elliptical}. In the bottom left plot, we have varied the maximum eccentricity values between 0 and 0.4 and in the bottom right, we expand the search to up-to a maximum eccentricity of 0.8. When more than one orbit is visible in the observation the templates clearly scale as $\sim \rm e_{max}^{2}$. We see the steepest dependence of templates on the maximum eccentricity parameter when 60 \% of the orbit is visible. In this case, for $\rm e_{max} = 0.4$, the templates scale approximately to the third power of eccentricity $\sim \rm e_{max}^{3}$ and for $\rm e_{max} = 0.8$, the templates scale $\sim \rm e_{max}^{6.5(4)}$. 

\section{Suitability for targeted and untargeted observations}
\label{sec: suitability}
In this section, we estimate the feasibility of applying a coherent search for circular and elliptical orbit binaries in untargeted pulsar surveys like the High Time Resolution Universe (HTRU; \citealt{2010MNRAS.409..619K}) as well as a targeted Globular Cluster (GC) observation of Terzan5. 

\subsection{HTRU-South Low Latitude Survey}

High Time Resolution Universe South Low-Latitude Survey (HTRU-S Lowlat) is one part of the HTRU Survey that focuses on the inner Galactic plane covering Galactic longitude \ang{-80} < \emph{l} < \ang{30} and Galactic latitude |\emph{b}| < \ang{3.5}. The observations were recorded with an integration time of 72 minutes with a bandwidth of 400~MHz split into 1024 channels. The data was recorded using the 21-cm multi-beam receiver along with the Berkeley–Parkes–Swinburne Recorder (BPSR) backend at the 64-m Parkes Radio Telescope. We refer the reader to \citet{2010MNRAS.409..619K} for a full list of the observational setup and system configuration. The main aim of this survey was to discover low-luminosity pulsars and new relativistic binary pulsars that are expected to be found close to the Galactic disk. HTRU-S Lowlat has 1230 pointings, with each pointing consisting of 13 beams. More than 100 new pulsars have already been discovered in HTRU-S Lowlat using a segmented acceleration search pipeline. We refer the readers to \citet{2015MNRAS.450.2922N} and \citet{2020MNRAS.493.1063C} for the initial list of discoveries and timing solutions. Given that we have already completed a first pass acceleration search over the entire dataset, this gives us a strong motivation to explore new techniques that can open up previously unexplored parameter spaces. We first start by estimating the computing time that would be required to search for non-recycled PSR-SBH binaries in HTRU using the techniques described in this paper. 
\subsection{Searching for Non-Recycled PSR-SBH Binaries in HTRU}
In the standard binary stellar evolution model \citep{1998astro.ph..1127Y, 2003MNRAS.342.1169V}, we expect the black hole to form first, followed by the pulsar at a later epoch. Therefore, in this formation channel we expect to find a slow spinning non-recycled pulsar orbiting a black hole in a wide eccentric orbit. Recent population synthesis work by \citep{2021MNRAS.504.3682C} has shown that MeerKAT is expected to find 0-30 PSR-BH systems and SKA is expected to observe 1–60 PSR+BHs in the Galactic field with most of them expected to contained unrecycled normal pulsars. A similar earlier study from \citet{2018MNRAS.477L.128S} has also estimated that there might be $\sim$3-80 PSR-BH binaries in the Galactic disk with about 10~\% that could be detected by Five-hundred-meter Aperture Spherical radio Telescope (FAST). In addition to the scaling relations described in section \ref{sec:scaling relation keplerian}, the number of orbital templates is also highly sensitive to the chosen spin frequency of the template bank. In circular orbit searches, templates scale as $f^3$ and in the elliptical orbit searches, templates scale as $f^5$. Therefore, a natural place to apply the template-bank algorithm is for the search of normal unrecycled pulsars in compact binary systems. 

For all our calculations in this section, we fix the spin period in our template bank at $P\mathrm{_{spin}} = 100$ ms. HTRU observations are natively sampled at \SI{64} {\micro \second}, however given that we are targeting non-recycled pulsars with a spin-period greater than or equal to \SI{100} {\milli \second}, it is safe to downsample the data by a factor of 16 to \SI{1024} {\micro \second}. This gives us access to up-to 32 harmonics of the pulsar signal below the Nyquist limit. Each observation is then dedispersed to dispersion measures of 0 to \SI{3000} {\parsec \per \centi \meter \cubed} typically consisting of 1126 DM trials assuming a DM smearing tolerance of 10\%.\footnote{The DM smearing tolerance value chosen for our benchmarks is based on the values used in the ongoing reprocessing of \textsc{HTRU-S Lowlat} by authors of this work. Results from these searches will be the subject of a future publication. The choice of the dedispersion step size is a function of the amount of computational resources available, the intrinsic spin-period and pulse-width of the pulsar as well as an acceptable pulse width broadening caused due to sampling at an incorrect DM. This adds an additional mismatch in practise for all binary search algorithms which is not taken into account in our tests in section \ref{sec:compare}. We refer the interested reader to section 6.1.1.2 of \citet{2012hpa..book.....L} for a more detailed discussion on this.} Using this and assuming we would fold at least 1000 candidates per beam, we estimate the total time it would take for our analysis in table \ref{tab:compute time young psr htru}. We calculate the cumulative time it would take to search using the random template-bank algorithm and folding on the full length HTRU observation as well as segmented searches on halves and quarter length observations. The total `Search' time includes applying a channel mask, Fourier birdie zapping, dedispersion, Fourier de-reddening, FFT, harmonic summing up-to 16 harmonics and a peak detection algorithm. Total time in shorter observations includes searching over all the relevant segments (e.g. 2 and 4 segments for 36 and 18 minute observations respectively). Our search performance was benchmarked on an Nvidia P100 GPU. 

The next step is called folding, where once the data is dedispersed, we sum the data at a particular spin-period based on the FFT period detection, orbital parameters and observation duration. This folded datacube is typically referred to as a pulsar candidate. We used the code \textsc{prepfold} from \textsc{PRESTO}  which also includes an optimization algorithm to find the best period and dispersion measure and is designed to be executed on a CPU. The folding performance was benchmarked on an Intel Xeon Gold 6140 processor. The total computation time reported in the table assumes the availability of 100 GPUs and 500 CPU cores. We added an overhead of 40\% for the total folding time to account for data transfer times, I/O, and non-linear scaling during parallel execution of folding. 

We fix the upper limit for the orbital period in our calculations to be $P\mathrm{_{orb}} = 10 T\mathrm{_{obs}}$ since an acceleration search would have already found binaries with orbital periods larger than this value. From the results presented in table \ref{tab:compute time young psr htru}, it is clear that for the vast majority of the parameter space, a five Keplerian parameter elliptical orbit binary search for non-recycled pulsars is not feasible for an untargeted pulsar survey like HTRU-S Lowlat. An exception is the regime between 2.5-10 $T\mathrm{_{obs}}$ which should be feasible to search with a maximum eccentricity of 0.1. For example, an elliptical orbit search for binaries between 1.5 to 6 hour orbital periods can be done in 145 days for the entire HTRU-S Lowlat survey searching across both half-length (36 minutes) observations. Any higher order eccentricity search is impractical with current and near-future computational resources. On a more optimistic note, \citet{2013MNRAS.432.1303B} showed that highly eccentric binaries ($e > 0.5$) are on average more favorable for detection compared to mildly eccentric systems ($e=0.1$) as the pulsar spends most of its time away from periastron and has a low line-of-sight acceleration for most of its orbit. Therefore, from both a physical and computational point of view, it is more favorable to stick to maximum eccentricity of 0.1. Additionally, from table \ref{tab:compute time young psr htru} we notice that a circular orbit binary search is computationally feasible and can open up interesting parameter spaces that are yet to be investigated. We notice that the searches between 1.7-10 $T\mathrm{_{obs}}$ can be analysed with current resources and even non-detections can be used to calculate limits on the existence of relativistic binary systems within our Galaxy. This is significant as a vast portion of these parameter spaces cannot be fully probed by existing polynomial-based searches to the same sensitivity provided by full orbital searches. For searches longer than 50 days, it may be desirable to reduce the total number of orbital templates (usually by a factor of 2) by applying the stochastic template-bank algorithm. This requires the investment of a few to 10 days worth of prior computing time using multiple cores on a HPC cluster in order to prune overlapping orbital templates. An implementation of this algorithm can also be found in our software repository. All our calculations in this and the following sections assume a coverage $\eta$ = 0.9 and a mismatch $m$ = 0.2. Therefore, our comments about feasibility of searches are based on these values. In principle, one can lower the coverage and increase the mismatch value in order to expand the search range of the orbital and spin parameters. The effectiveness of this approach compared to existing pipelines need to be tested and discussions related to this is beyond the scope of this paper.

\begin{table*}
\caption{Total Computing time required to search for non-recycled PSR-SBH binaries ($P\mathrm{_{spin, min}} = 100$ ms.) using a coherent circular binary and elliptical binary search in the entire HTRU-S Lowlat Survey. Total 'Search' time includes applying a channel mask, Fourier birdie zapping, dedispersion, Fourier de-reddening, FFT, harmonic summing up-to 16 harmonics and a peak detection algorithm. Our searches were benchmarked on a Nvidia P100 GPU and the folding was benchmarked on an Intel Xeon Gold 6140 processor using \textsc{prepfold} from \textsc{PRESTO} .}
\label{tab:compute time young psr htru}
\begin{center}
\begin{threeparttable}
\begin{tabular}{lrrrrrrrrrr}
\toprule
$t_\text{obs}$ & Min. $P_\text{b}$ & Max. $P_\text{b}$ &  Mass$\mathrm{_{companion}}$ &  Max. Eccentricity\tnote{a}  & \multicolumn{3}{c}{Circular Binary Search} & \multicolumn{3}{c}{Elliptical Binary Search} \\
\cmidrule(lr){6-8}\cmidrule(lr){9-11}
(mins) & (h) & (h) & ($M_{\odot}$) &  & Search(d) & Folding(d) & Total (days) & Search(d)  & Folding(d) & Total (years) \\
\midrule
72 & 3 & 12 & 8 & 0.1 &16.2 & 35.4 & 51.6 & 1613.6 & 35.4 &  4.5  \\
72 & 2 & 12 & 8 & 0.1 & 229.6 & 35.4 & 265 & 660,079.5 & 35.4 & 1808  \\
36 & 1.5 & 6 & 8 & 0.1 & 5 & 19 & 24 & 145 & 19 & 0.45 \\
36 & 1.0 & 6 & 8 & 0.1 & 68.7 & 19 & 87.7 & 59,717.8 & 19 & 163.7  \\
18 & 0.42 & 3 & 8 & 0.1 & 70.8 & 10 & 80.8 & 108,780.9 & 10 & 298  \\
\bottomrule
\end{tabular}
\begin{tablenotes}
  \item[a] Applicable only for Elliptical Orbit Searches.
  \end{tablenotes}
  \end{threeparttable}
\end{center}
\end{table*}

\subsection{Searching for Millisecond PSR-SBH Binaries in HTRU}
The formation of a millisecond pulsar (MSP) orbiting a black hole is expected to be rare. One possible scenario is the so-called reversal mechanism where under certain conditions the pulsar is formed first and is later spun-up by accretion during the red giant phase of the secondary star \citep{2004MNRAS.354L..49S, 2005ApJ...628..343P}.
Discovering a MSP-Black Hole binary is more desirable as MSPs tend to be more precise timers than slow `normal' pulsars \citep{2009MNRAS.400..951V}. Another possibility is the dynamical formation scenario due to exchange interactions. This is more likely in regions of high stellar density like the Galactic centre. While interstellar scattering is expected to hamper detections of PSR-BH binaries near Sgr A* at L-band in HTRU-S Lowlat, it is still important to search for such systems due to their potential scientific impact. Given that elliptical binary searches were already unfeasible for searching for unrecycled PSR-SBH binaries, the situation is expected to be worse for recycled PSR-SBH binaries. However, circular binary searches are still feasible and we provide a few examples of search-setups in table \ref{tab:compute time msp psr htru}. We focus mainly on the spin-period regime between 13-20 ms. MSPs rotating faster than this are still computationally unfeasible to search for PSR-SBH binaries. An alternate possibility is to reduce the companion mass limit and focus searches for double neutron-star and pulsar-white dwarf binaries. This would help in reducing the spin-period limit and search for more recycled pulsars. \citet{2011PhDT.......293K, 2013ApJ...773...91A} and \citet{2013ApJ...774...93K} used this set-up to search for pulsars in PALFA \citep{2006ApJ...637..446C} and PMPS \citep{2001MNRAS.328...17M}) respectively using the volunteer distributed computing resources available from the Einstein@Home project. Our calculations indicate that the regime between 5-10 $T\mathrm{_{obs}}$ is feasible to search for MSP-SBH binaries in HTRU with spin-periods greater than 13 ms using a circular binary search. The entire analysis would take anywhere between 6-18 months depending on the chosen search setup. Additionally, having a higher spin-period threshold at 20 ms could help us investigate binaries with orbital period in the  3-10 $T\mathrm{_{obs}}$ regime. Our benchmarks were done assuming the same computational resources as in the previous section. In this case, we downsampled the data to \SI{256}{\micro \second} which gives us access to at least 16 harmonics of a hypothetical 13 ms pulsar signal. The total search time includes 1876 dedispersion trials assuming a DM smearing tolerance of 10\% from 0 to \SI{3000} {\parsec \per \centi \meter \cubed}.

\begin{table*}
\caption{Total Computing time required to search for MSP PSR-SBH binaries using a coherent circular binary search in the entire HTRU-S Lowlat Survey. For searches longer than 50 days, we recommend applying the stochastic template-bank algorithm and pruning overlapping templates. This usually reduces the number of templates by a factor of 2 at the cost of approximately an extra week of up-front computing time.}
\label{tab:compute time msp psr htru}
\begin{center}
\begin{threeparttable}
\begin{tabular}{lrrrrrrr}
\toprule
$t_\text{obs}$ & Min. $P_\text{b}$ & Max. $P_\text{b}$ &  Mass$\mathrm{_{companion}}$ &  P$\mathrm{_{spin}}$  & \multicolumn{3}{c}{Circular Binary Search} \\
\cmidrule(lr){6-8}
(mins) & (h) & (h) & ($M_{\odot}$) & (ms) & Search(d) & Folding(d) & Total (days)  \\
\midrule
72 & 6 & 12 & 8 & 13 & 355.6
& 214.0 & 569.6  \\
72 & 6 & 12 & 8 & 20 & 99.0 & 214.0  & 313.0  \\
36 & 3 & 6 & 8 & 20 & 99.0 & 86.5 & 185.5  \\
36 & 2 & 6 & 8 & 20 & 400.3 & 86.5 & 486.8  \\
18 & 1 & 3 & 8 & 20 & 120.0 & 67.3 & 187.3 \\
9& 0.42 & 1.5 & 8 & 20 & 99.3 & 58.0 & 157.3 \\
\bottomrule
\end{tabular}

  \end{threeparttable}
\end{center}
\end{table*}

\subsection{Searching for MSP-SBH binaries in Globular Clusters}
Globular Clusters (GCs) are one of the ideal locations to apply the template-bank algorithm. GCs produce orders of magnitude more MSPs per unit
mass than the Galactic disk and since most MSPs are expected to be binaries, we expect to find several compact binary pulsars in GCs. There are currently 233 known pulsars in 36 globular clusters\footnote{Refer to \url{http://www.naic.edu/~pfreire/GCpsr.html} for an updated list of the catalog.} Additionally, the advantage of processing GCs is that the DM of the cluster maybe well known which means most of the compute power can be invested in exploring the binary parameter phase space. Any new candidate can also be confirmed potentially using archival observations of these clusters. Therefore, we could have a timing solution that spans across several years within a short period of time after detection, thus enabling us to get scientific results quicker. 

In the context of PSR-SBH binaries, analogous to the Galactic centre, the high stellar density in GCs opens up the possibility of three-body interactions where a neutron star can gain a companion by exchanging with a primordial binary and subsequently be spun up to become an MSP \citep{1976MNRAS.175P...1H, 1995ApJS...99..609S, 2008MNRAS.386..553I, 2014MNRAS.442..207C}. Another possibility is the formation a MSP in orbit with an intermediate-mass black hole (IMBH: $M \sim 10^2$ - $10^4 M_{\odot}$). The encounters between MSPs and IMBHs could result in the MSP being significantly displaced from the core \citep{2003ApJ...599.1260C} to form a MSP-IMBH binary \citep{2007MNRAS.380..691D}. Recent theoretical work using N-body simulations by \citet{2019ApJ...877..122Y} has shown that the number of MSPs and BHs in GCs
are likely anti-correlated. Therefore, the chances to find an MSP-SBH binary are probably lower than previously assumed. Nevertheless, this question can only be fully addressed by doing searches on real data using different binary search techniques. Here, we investigate the compute time required to search for such binaries using both a circular and elliptical orbit search in table \ref{tab:compute time GCs}. We used a 7.2 hour archival observation of Terzan5 taken at the 100-m Robert C. Byrd Green Bank Telescope (GBT) for our benchmarks \citep{2017ApJ...845..148P}. The observation was recorded at a central frequency of 2000 MHz (S-band) using a bandwidth of 800 MHz split into 128 channels. The data recorded by the Green Bank Ultimate Pulsar Processing Instrument (GUPPI) were full-Stokes with a sampling time of \SI{10.24}{\micro \second} and 512 channels, each coherently dedispersed to a DM of \SI{238} {\parsec \per \centi \meter \cubed}. 

For our benchmarks, we downsampled the data to \SI{81.9}{\micro \second} and assumed 50 dedispersion trials and the same computational resources as the previous section. For the elliptical searches, we restrict the maximum eccentricity to 0.1. Any higher order eccentricity search is still not computationally feasible. As expected processing longer integration times are computationally more challenging. Based, on our benchmarks a circular-orbit search between 5-10 $T\mathrm{_{obs}}$, i.e.\ 36-72 hour orbital period binaries for a 7.2-hour observation up to a companion mass of 10 $M\mathrm{_{\odot}}$, would take about 1.3 days assuming 100 GPUs for searching and 500 CPU cores for folding. We also provide rough estimates for segmented searches. The total processing time for observations shorter than 7.2 hours includes searching all search segments (for e.g. 2 segments for a 3.6-hour observation). Elliptical orbit binary searches are not feasible for very long integration times. In order to make the elliptical binary search feasible, we need to either search on shorter length observations or have a higher threshold for the spin-period in our template-bank ($P{_\mathrm{{spin}}} \geq \SI{40} {\milli \second}$). We have given a few examples of possible parameter-space combinations that can be explored with current computational resources. This is given in the right hand side of table \ref{tab:compute time GCs}. Some examples readers might find interesting include an elliptical orbit binary search on a 1.8 hour observation with a spin-period threshold of 40 ms and a maximum companion mass of 100 $M\mathrm{_{\odot}}$. Reducing the coverage of the template-bank to 0.7 and increasing the mismatch to a value of 0.5 can make a search with minimum spin-period threshold of 20 ms feasible. Additionally, we can go down to a minimum spin period of 5 ms for an elliptical orbit search for a 30 minute search observation. Another interesting example would be a circular orbit binary search between 5-10 T$_{\rm obs}$ orbital period regime with a maximum companion mass of 1000 $M\mathrm{_{\odot}}$. This is possible for observation times lesser than or equal to an hour. Searching for binary pulsars with a spin-period of 1 ms and a maximum companion mass of 1000 $M\mathrm{_{\odot}}$ is also feasible for observations of integration length 30 minutes or lower. Yet another possibility to reduce computation time is to focus on double neutron star searches. This would help to investigate faster spinning pulsars in more compact orbits. For example, using a circular binary search for a 30 minute observation, we could search all 14 segments (total length = 7.0 hrs), with an orbital period range of 1-5 hours (2-10 $T\mathrm{_{obs}}$) with a maximum companion mass of 1.6 $M\mathrm{_{\odot}}$ in a day.

\begin{table*}
\caption{Total Computing time required to search for PSR-SBH binaries using a coherent circular binary and elliptical binary search in a GBT observation of Terzan5. Duration includes time taken for searching and folding. These numbers assume 100 Nvidia P100 GPUs and 500 CPU cores. The total time taken for observation shorter than 7.2 hours includes time taken to cover all search segments. }
\label{tab:compute time GCs}
\begin{center}
\begin{threeparttable}
\begin{tabular}{lrrrrrrrrr}
\toprule
$t_\text{obs}$ & Min. $P_\text{b}$ & Max. $P_\text{b}$   &     \multicolumn{3}{c}{Circular Binary Search} & \multicolumn{4}{c}{Elliptical Binary Search} \\
\cmidrule(lr){4-6}\cmidrule(lr){7-10}
(hrs) & (h) & (h)  &  P$\mathrm{_{spin}}$ (ms) & Mass$\mathrm{_{companion}}$ ($M_{\odot}$) & Duration (days)\tnote{a}  &  P$\mathrm{_{spin}}$ (ms) & Mass$\mathrm{_{companion}}$ ($M_{\odot}$) & Max. Eccentricity\tnote{a}  &  Duration (days)\tnote{a} \\
\midrule
7.2 & 36 & 72 & 10 & 10 & 1.3 & 40 & 10 & 0.1 & 27.1  \\
3.6 & 18 & 36 & 10 & 10 &  0.24 & 20\tnote{b} & 10 & 0.1 & 3.7  \\
1.8 & 9 & 18 & 10 & 10 &  0.06 & 40 & 100 & 0.1 & 2.2  \\
1.0 & 5 & 10 & 10 & 1000 &  2.0 & 15\tnote{b} & 100 & 0.1 & 1.84  \\
0.5 & 2.5 & 5 & 1 & 1000 & 0.14 & 5 & 10 & 0.1 & 23.74 \\
0.5 & 1.5 & 5 & 10 & 100 & 1.69 & 20 & 10 & 0.1 & 2.56 \\
0.5 & 1.0 & 5 & 5 & 1.6 & 0.9 & 40 & 1.6 & 0.1 & 0.9 \\

\bottomrule
\end{tabular}
\begin{tablenotes}
  \item[a] Includes time taken for search and folding.
  \item[b] This search set-up assumes a coverage $\eta = 0.7$ and mismatch $m = 0.5$.
  \end{tablenotes}
  \end{threeparttable}
\end{center}
\end{table*}

\section{Discussions and Conclusions}
A question not addressed in our work is the special case of targeted searches i.e.\ when one or multiple Keplerian parameters are known accurately enough that a search across that dimension is not needed. We refer the readers to section 3-B of \citep{2001PhRvD..63l2001D} for a more detailed formalism of creating a template-bank in this scenario. This can significantly reduce the number of orbital templates required for searching. We expect this to be incorporated in a future version of the software. Another promising approach that requires further investigation in the context of radio pulsar searching is hierarchical searches. \citet{2020ApJ...901..156N} described a multi-stage search where in the first stage the parameter space is investigated at a lower sensitivity (high mismatch), followed by searching in smaller regions of the parameter space around promising candidates with higher sensitivity. This technique was successfully used to discover a binary Gamma-Ray Pulsar in Fermi Large Area Telescope (LAT) data \citep{2020ApJ...902L..46N}. In this paper, we described the implementation for creating a template-bank to do a fully coherent search across three and five Keplerian parameters assuming circular and elliptical orbits respectively. We demonstrated the extra sensitivity gained by applying template-bank pipeline compared to acceleration and jerk searches from \textsc{PRESTO} and \textsc{sigproc} for simulated Pulsar stellar-mass Black hole binaries and the double pulsar \mbox{PSR~J0737-3039A}. We have also given rough scaling relations for the total required orbital templates with respect to observation time, minimum orbital period, projected radius and maximum companion mass for both circular and elliptical orbit searches in addition to maximum eccentricity for elliptical orbit searches. Trade-offs of this algorithm include a significantly longer computation time and reduced sensitivity to longer orbital period binaries ($P_{\rm orb} > 10 T$) due to them being outside our search range and the increased number of orbital trials. Another limitation is our reduced sensitivity to isolated fast-spinning pulsars due to our high spin-period threshold which would be necessary in order to make the search computationally feasible for PSR-SBH binaries. We also benchmarked the amount of time it would take to search for non-recycled PSR-SBH binaries in HTRU-S Lowlat. We note that the regime between 2.5-10 $T\mathrm{_{obs}}$ should be feasible to search in 36-minute half length observations of HTRU-S Lowlat with a maximum eccentricity of 0.1 and a spin-period threshold of 100 ms. Searching across all five Keplerian parameters for millisecond PSR-SBH binaries on a survey like HTRU-S Lowlat is still not computationally feasible. However, circular binary searches could open up interesting parameter spaces here that are yet to be explored by acceleration and jerk searches. For example, assuming a maximum companion mass of 8 $M\mathrm{_{\odot}}$, the regime between 5-10 $T\mathrm{_{obs}}$ should be feasible to search in full length HTRU-S Lowlat observations with a spin-period greater than or equal to 13 ms. By reducing the spin-period threshold to 20 ms, we could explore binaries with orbital period in the 3-10 $T\mathrm{_{obs}}$ regime. We expect the most applicability of the template-bank pipeline on targeted observations like globular clusters. Here, our benchmarks indicate that searching for mildly recycled pulsars orbiting an intermediate mass black hole is feasible for observations shorter than 2 hours with a maximum eccentricity of 0.1 in the template-bank. Our paper focused more on the application of the template-bank pipeline for PSR-SBH binaries. However, several other search strategies are also possible using this approach targeting PSR-WD binaries or DNS systems and we hope going forward this will be a useful and complementary approach to the traditionally used acceleration and jerk search pipelines in literature. In order to aid future binary pulsar searches using the template-bank pipeline, we additionally also open-source our circular\footnote{\url{https://github.com/vishnubk/3D_peasoup}} and elliptical\footnote{\url{https://github.com/vishnubk/5D_Peasoup}} binary search GPU pipelines.

\section*{Acknowledgements}
We would like to thank Norbert Wex and Benjamin Knispel for useful discussions. We would also like to thank Scott Ransom for providing us with the GBT observation of Terzan5. The data analysis were performed on the OzSTAR national supercomputing facilities at Swinburne University of Technology and the HERCULES computing cluster operated by the Max Planck Computing \& Data Facility (MPCDF). OzSTAR is funded under Astronomy National Collaborative Research Infrastructure Strategy (NCRIS) Program via Astronomy Australia Ltd (AAL). We would like to thank members of the open-source community for maintaining software packages that were directly used for our work including \textsc{NumPy} \citep{5136190}, \textsc{Matplotlib} \citep{Hunter:2007}, \textsc{emcee} \citep{emcee}, \textsc{corner} \citep{corner}, \textsc{SymPy} \citep{10.7717/peerj-cs.103} and \textsc{CUDA} \citep{cuda}.

\section*{Data Availability}
The data underlying this article will be shared on reasonable request to the corresponding author.



\bibliographystyle{mnras}
\bibliography{latest_draft} 




\appendix
\section{Software and Implementation Details}
\label{sec:software}
Our search code has been designed to run on Nvidia-GPUs. It is built on top of the GPU-accelerated \textsc{PEASOUP}\footnote{\url{https://github.com/ewanbarr/peasoup/}} pipeline that does an acceleration search \citep{1991ApJ...368..504J}. The main pipeline also implements dedispersion through the \textsc{DEDISP} library \citep{2012MNRAS.422..379B}, Fourier de-reddenning, FFT, incoherent harmonic summing and a peak detection algorithm for candidate selection. Our version of the software called \textsc{5D-Peasoup}\footnote{\url{https://github.com/vishnubk/5D_Peasoup}} replaces the acceleration search with a coherent search across all Keplerian parameters and additionally runs a candidate reduction algorithm by grouping together multiple detections of the same candidate.
\subsection{Time-Domain Resampling Algorithm}
\label{sec:time domain resampling}
We apply the nearest-neighbour resampling algorithm which is the standard resampling technique applied in most time-domain pulsar search pipelines including \textsc{SIGPROC}, \textsc{PEASOUP} and the Einsten@Home Radio Pulsar-Search pipeline \citep{2011PhDT.......293K,2013ApJ...773...91A}. The difference here is that our Roemer delay equation contains all five Keplerian parameters. Our goal is to transform our reference frame from the observed timeseries at the telescope to the binary system's barycenter. After correcting for the dispersion delay caused by free electrons in the Inter-Stellar Medium (ISM) using the software \textsc{DEDISP}, we additionally need to account for the Roemer delay caused due to the motion of the pulsar around its companion \citep{1976ApJ...205..580B}. For a Keplerian orbit with non-zero eccentricity, this can be written as:
\begin{equation}
  \Delta_R = \tau [(\cos E - e) \sin \psi + \sqrt{
  1 - e^2} \sin E \cos \psi].
  \label{eq: romer delay bt}
\end{equation}

As mentioned earlier calculating the sine and cosine of eccentric anomaly involves solving Kepler's equation iteratively. The alternative is to expand $\cos E$ and $\sin E$ as a power series shown in equation \ref{eq: cosE ell}, \ref{eq: sinE ell}. The order to which the power series should be expanded depends on the eccentricity range and the companion mass we want to be sensitive to. These models are referred to as ELL1 (Wex 1998, unpublished contribution to \textsc{TEMPO}, see also \citealt{2001MNRAS.326..274L}), ELL2 and so on based on the number of retained terms in the power series. For the calculation of the template-bank, we used the ELL7 model which covers up to $e\approx 0.8$. However, in our resampling algorithm we implemented the BT model. Therefore, the time $t^{\prime}$ at the barycenter can be calculated by solving 

\begin{equation}
t^{\prime} = t - \tau [(\cos E - e) \sin \psi + \sqrt{
  1 - e^2} \sin E \cos \psi],
\end{equation}
where we have ignored constant phase shifts caused by initial signal phase and the constant classical light travel time between the pulsar and the barycenter.
The signal $S$ is recorded at the telescope in discrete time steps $S = i \times t\mathrm{_{samp}}$ where i is the bin number and $t\mathrm{_{samp}}$ is the sampling time and our goal is to calculate $S^{\prime} = j \times t\mathrm{_{samp}}$ which is the signal at the barycenter.
We first start by calculating a zero offset ($t = 0$) for each orbital template, in order to shift the scale of the observed timeseries to match with the barycentered timeseries. This number is then subtracted from the Roemer delay, which is usually a rational number and then we calculate the nearest integer to this number and map the value in that bin to our resampled timeseries. A brief description of this can be found in algorithm \ref{alg:resampling}.

\begin{algorithm}
\SetAlgoLined
\DontPrintSemicolon
\SetAlgoRefName{1}
\caption{Nearest-Neighbour Demodulation Algorithm}

\label{alg:resampling}
\For{\textsf{\upshape K orbital templates in template bank}}{

Calculate Mean Anomaly $M$ at time $t = 0$
\begin{equation*}
M = \alpha
\end{equation*}
Solve for Eccentric Anomaly $E$ iteratively
\begin{equation*}
E - e \sin E = M,
\end{equation*}
Compute Zero-Offset ${t_0}$ \;
 \begin{equation*}
  \mathrm{t_0} =  \frac{\tau [(\cos E - e) \sin \psi + \sqrt{
  1 - e^2} \sin E \cos \psi]}{t\mathrm{_{samp}}}
\end{equation*}
\For{\textsf{\upshape $i$th bin in observed timeseries}}{
\begin{equation*}
M = \Omega ((i \times t\mathrm{_{samp}}) + \alpha)
\end{equation*}
Solve for $E$ iteratively
\begin{equation*}
E - e \sin E = M,
\end{equation*}
\begin{equation*}
S^{\prime}_j = S\left [j - \frac{\tau [(\cos E - e) \sin \psi + \sqrt{
  1 - e^2} \sin E \cos \psi]}{t\mathrm{_{samp}}} - \mathrm{t_0} \right ]_{\rm int}
\end{equation*}

}
return $S^{\prime}$ to calculate FFT

}
\end{algorithm}
\section{Coefficients of the taylor expansion of Eccentric Anomaly}
\label{sec: coefficients}
Below, for completeness we mention the coefficients C$_k$, S$_k$ taken from \citep{1985cmcg.book.....T} and used in equations \ref{eq: cosE ell} and \ref{eq: sinE ell}. These expressions were also given in Appendix A of \citet{2001PhRvD..63l2001D}. A derivation for these expressions can also be found in Appendix C of \citep{2020ApJ...901..156N}. 
\begin{align*}
C_0 &= -\frac{1}{2} e, \\
C_1 &= 1 - \frac{3}{8}e^2 + \frac{5}{192} e^4 - \frac{7}{9216}e^6, \\
C_2 &= \frac{1}{2}e - \frac{1}{3}e^3 + \frac{1}{16}e^5, \\
C_3 &= \frac{3}{8}e^2 - \frac{45}{128}e^4 + \frac{567}{5120}e^6, \\
C_4 &= \frac{1}{3}e^3 - \frac{2}{5}e^5,\\
C_5 &= \frac{125}{384}e^4 - \frac{4375}{9216}e^6,\\
C_6 &= \frac{27}{80}e^5, \\
C_7 &= \frac{16807}{46080}e^6,
\end{align*}
\begin{align*}
S_1 &= 1 - \frac{5}{8}e^2 - \frac{11}{192}e^4 - \frac{457}{9216}e^6, \\
S_2 &= \frac{1}{2}e - \frac{5}{12}e^3 + \frac{1}{24}e^5, \\
S_3 &= \frac{3}{8}e^2 - \frac{51}{128}e^4 + \frac{543}{5120}e^6, \\
S_4 &= \frac{1}{3}e^3 - \frac{13}{30}e^5, \\
S_5 &= \frac{125}{384}e^4 - \frac{4625}{9216}e^6, \\
S_6 &= \frac{27}{80}e^5, \\
S_7 &= \frac{16807}{46080}e^6.
\end{align*}


\bsp	
\label{lastpage}
\end{document}